\documentclass[graybox]{svmult}

\usepackage{type1cm}        % activate if the above 3 fonts are
\usepackage{makeidx}         % allows index generation
\usepackage{graphicx}        % standard LaTeX graphics tool
\usepackage{multicol}        % used for the two-column index
\usepackage[bottom]{footmisc}% places footnotes at page bottom

\usepackage{newtxtext}       % 
\usepackage{newtxmath}       % selects Times Roman as basic font

\makeindex             % used for the subject index
\usepackage{amsmath}

\begin{document}
%\tableofcontents{}
\title*{Time domain methods for X-ray and gamma-ray astronomy}
\author{Eric D. Feigelson\thanks{corresponding author}, Vinay L. Kashyap and Aneta Siemiginowska}

\institute{Eric D. Feigelson \at Center for Astrostatistics, Department of Astronomy \& Astrophysics, Penn State University, University Park PA 16802 USA \email{e5f@psu.edu} \and
Vinay L. Kashyap \at Harvard/Smithsonian Center for Astrostatistics, 60 Garden Street, Cambridge MA 02138 USA \email{vkashyap@cfa.harvard.edu}
\and 
Aneta Siemiginowska \at Harvard/Smithsonian Center for Astrostatistics, 60 Garden Street, Cambridge MA 02138 USA \email{asiemiginowska@cfa.harvard.edu}}

\maketitle

\abstract{
A variety of statistical methods for understanding variability in the time domain for low count rate X-ray and gamma-ray sources are explored.  Variability can be detected using nonparametric (Anderson-Darling and overdispersion tests) and parametric (sequential likelihood-based tests) tools.  Once detected, variability can be characterized by nonparametric (autocorrelation function, structure function, wavelet analysis) and parametric (multiple change point model such as Bayesian Blocks, integer autoregressive models, C-statistic and Poisson regression) methods. New multidimensional variability detection approaches are outlined. Software packages designed for high energy data analysis are deficient but tools are available in the R statistical software environment. Most of the methods presented here are not commonly used in high energy astronomy. 
}

~\footnote{
To appear in Handbook for X-ray and Gamma-Ray Astrophysics, 
Volume 4: Analysis techniques, 
Section XVIII: Timing Analysis  (Belloni \& Bhattacharya, eds.)
} \\~\\

\section{Variability in High Energy Astronomy}
\label{intro.sec}

The X-ray and gamma-ray sky is dominated by variable sources.  The brightest sources are often close binary star systems in the Milky Way Galaxy where gas accretes from a normal companion onto a neutron star or black hole.  Here periodic components can include slow orbital eclipses and rapid rotation of the neutron star.  But stochastic effects are also present with both random and structured variations such as quasi-periodic oscillations arising from processes within the accretion disk.  Other classes of variable high energy emitters include magnetic flares from normal stars, thermonuclear explosions on white dwarfs and neutron stars, supernovae from all types of exploding stars, quasars (where the X-ray emission arises from a hot corona or irradiated accretion disk) and blazars (where the emission arises from a nonthermal jet) produced by accreting supermassive black holes.  Even extended supernova remnants and radio galaxy jet knots exhibit slow, barely perceptible changes. 

Four basic challenges of temporal analysis in high energy astronomy can be identified. Considered in various combinations for a given problem, a rigorous and reliable statistical and scientific analysis of variable X-ray and gamma-ray sources can be a complex enterprise. 
\begin{enumerate}
    \item Variations occur on an enormous range of timescales from milliseconds to decades.  Different instruments are needed to capture variations on different timescales: high-throughput devices like the NICER X-ray Timing Instrument for the most rapid variations; long-lived all-sky monitors for the longest timescales; and telescopes like the Chandra X-ray Observatory, XMM-Newton and the Neil Gehrels Swift Observatory for intermediate timescales. 
    
    \item Brightness changes exhibit a wide range of behaviors that can be deterministic or stochastic, aperiodic or periodic.  Rotational or orbital effects are predictable except for minute period changes of astrophysical interest.  Accretion disk emission and hot spots on the white dwarf or neutron star can exhibit complex phenomena; for example, the X-ray binary Sco X-1 has red noise and quasi-periodic oscillations that can be represented by transient chaos on a dripping rail \cite{Scargle93}. Even a single source can exhibit a bewildering variety of phenomena: the black hole binary GRS 1915+105 switches between a dozen states, due perhaps to complex disk-jet interactions. \cite{Fender04, Huppenkothen17}. Quasar variations are simpler; their variations can often be modeled as a low-dimensional linear autoregressive process \cite{Kelly09,Kelly2011}.
    
    \item The datasets can have some unusual characteristics such as irregular observations with gaps and multiple dimensions.  Gaps in the data stream can arise every $\sim 90$ minutes for satellites in low-Earth orbit, and often are produced by non-optimal telescope allocation scheduling.  Instrumental issues can also affect temporal characteristics such as detector dead-time and pileup for bright sources. 
    
    \item High energy light curves are often treated as a multivariate process.  The most common multivariate studies concern how brightness variations relate to the source spectra.  Plots of brightness against hardness ratios, plasma temperature (for thermal sources) or spectral index (for nonthermal sources), or multiband observations can give powerful insights into astrophysical processes.  The evaluation of ratios of low-count integers is a surprisingly tricky statistical problem but effective approaches are available \cite{Brown01, Park06}. 
\end{enumerate}

The statistical foundation of X-ray and gamma-ray data can be Gaussian or Poissonian depending on the count rate.  As time series analysis of Gaussian data can be based on familiar methods, most of this review is limited to treatments of low-count rate data modeled with a Poisson distribution.

\section{Methodological Foundations for High Energy Light Curves}
\label{foundations.sec}

The underlying structure of high energy astronomical datasets is typically a 4-dimensional multivariate stochastic process.  It consists of a list of photons each with celestial location (two dimensions), energy in keV or MeV, and arrival time.  Commonly, the astronomer analyzes the dataset sequentially: identifying clusters of photons in celestial location for source detection and photon extraction; analysis of the energy distribution to understand the physical processes of emission; and analysis of the photon arrival times to assess whether the source is variable. Thus, univariate time series are most commonly treated.   But it is possible have multivariate time series by dividing the photons into bands, or to consider the dataset in its original 4-dimensional form (\S\ref{4d.sec}).

Four types of datasets are encountered in X-ray and gamma-ray astronomy:
\begin{description}
    \item[\bf Gaussian, regularly spaced bins]  When sufficiently large numbers of photons are encountered that each observation interval is well-populated (say, $>20$ events), and the intervals are regularly spaced with few or no interruptions, then standard time series analysis for real-valued data can be pursued. Widely used methods such as the $\chi^2$ and excess variance tests for variability, smoothing procedures, and least-squares fitting of autoregressive or flare models can be applied. More thorough methodological treatments are provided in papers such as Scargle (1981), Koen \& Lombard (1993), and Vaughan et al. (2003). An example of an analysis of this type appears in the textbook by Feigelson \& Babu (2012) with cookbook for R code.  Here the dataset consists of $\sim 2$ million photons detected with the $Ginga$ satellite at high time resolution over $\sim 2$ hours from the Galactic quasi-periodic oscillator source GX~5-1.  This case is not treated in this chapter.
 
    \item[\bf Gaussian, irregularly spaced observations]  Here the observations are collected into real-valued measurements of flux for a number of epochs that may have uneven spacing due to satellite constraints or telescope allocation times.  Analysis methods used by optical astronomers can be used for such datasets such as: mean, standard deviation, and chi-squared; medians, InterQuartile Range, and quantiles; Welch-Stetson indices; Edelson-Krolik discrete correlation function; and more.  Reviews of these methods can be found in Bellm (2021) who considers metrics for variability analysis for the Rubin Observatory Legacy Survey of Space and Time.  This case is also omitted in this chapter. 
 
    \item[\bf Poissonian, regularly spaced bins]  Here the instrument or scientist collects the X-ray or gamma-ray events into regularly spaced time intervals, each of which has integer-valued counts of a few (roughly $<20$) photons so that the Gaussian approximation is no longer accurate.  Some bins may have zero events.  In statistical parlance, such time series are examples of {\bf count data}.   We treat below analysis of such datasets both ignoring the temporal sequence and for regularly spaced temporal bins (\S\ref{overdispersion.sec}-\ref{other_nonpar.sec}, \S\ref{characterization.sec}).

    \item[\bf Poissonian, irregularly spaced individual events]  Finally we consider the data as it generally arrives from the instrument, not as measurements of count rates or fluxes, but rather as a sequence of event arrival times.  Each event may be tagged with estimates of location and energy.  The analysis methodology for this case is quite different than the others, arising more from the field of stochastic processes than from classical statistics.  We present a variety of methods for this case (\S\ref{AD.sec}, \S\ref{other_nonpar.sec}, \S\ref{changepoint.sec}, \S\ref{astrophys.sec}, \S\ref{4d.sec}). 
\end{description}

Some basic concepts of time series are needed to select appropriate methods for analysis.  A process giving rise to a time series is {\bf stationary} if its temporal behavior has the same properties for any time interval.  The simplest case is a source with constant flux, exhibiting Gaussian or Poissonian white noise.  But a source that is periodic with constant period and phase, or autocorrelated with constant autoregressive coefficients (think of a bubbling brook), can also be stationary.  The widely used `damped random walk' model for a quasar with constant fluctuation properties is an example of  parametric autoregressive modeling of a stationary process \cite{Kelly14}.  In such cases, methods that treat the dataset in a global fashion $-$ such as moments (e.g. mean and standard deviation), the chi-squared test, autocorrelation function,  autoregressive modeling \cite{Box15, Chatfield19}, and Fourier analysis \cite{Priestley81} $-$ are appropriate. Fourier and other frequency domain methods for uncovering periodic behaviors are presented in other chapters.    

A {\bf nonstationary} process has temporal behaviors that are not constant throughout the observation.  The simplest cases involves changes in mean flux, ranging from gradual brightening to sudden outbursts.  But there could also be changes in standard deviation (called volatility) in autoregressive or periodic behaviors. Nonstationarity can arise from the intermingling of two or more signals. Note that a short duration time series might appear as nonstationary even though the long term behavior is stationary.  A useful review of statistical approaches to nonstationary time series is provided by Salles et al. (2019). 

In high energy astronomy, Vaughan et al. (2003) show how nonstationary long-memory autoregressive behaviors, commonly called 'red noise', affects the variance of binned light curves.  Some high energy light curves are spectacularly nonstationary such as prompt gamma-ray bursts and accretion systems like GRS 1915+105 with dozens of variability states  \cite{Fender04, Huppenkothen17}.  In such cases, local regression, wavelet analysis \cite{Mallat08, Nason08} and multiple change point analysis \cite{Tartakovsky20} might be effective analysis methods.  The Bayesian Blocks decomposition popular in high energy astronomy is an example of a likelihood-based parametric change point modeling procedure for nonstationary high energy lightcurves \cite{Scargle13} (\S\ref{changepoint.sec}).  

It is also important to distinguish {\bf parametric} and {\bf nonparametric} approaches to time series analysis.  Consider smoothing, where the scientist seeks to convert measurements, either individual or binned photons, into a continuous curve of source brightness as a function of time.  Nonparametric approaches include procedures like running medians, kernel density estimation (e.g. convolving the flux light curve with a Gaussian), and local regression.  Local nonparametric  regression itself has a variety of approaches \cite{Takezawa06, Harezlak18}: spline fits (since the 1950s), Nadaraya-Watson kernel regression (since the 1960s), local polynomial fits (such as the LOESS model in the 1970s), and  Gaussian Processes regression (known as 'kriging' since the 1960s).   {\bf Semi-parametric} regressions (splines, kernel-based regressions and Bayesian Blocks) have some local parametric assumptions but no global parametric model.  Smoothing can be used both to examine the evolution of source brightness, and as a model to subtract from the data to study any residual fluctuations.  

Formal statistical hypothesis tests can be applied to regularly spaced light curves \cite{Enders14}.  For Gaussian processes, there are tests for normality (e.g. Shapiro-Wilk and Anderson-Darling tests), stationarity (adjusted Dickey-Fuller and Kwiatkowski-Phillips-Schmidt-Shin tests), and  autocorrelation (Durbin-Watson, Brosch-Pagan, and Ljung-Box tests).  We discuss below tests for a homogeneous Poisson distribution (\S\ref{detection.sec}).  Hypothesis tests can be applied both to observed light curves and to residuals after subtracting a nonparametric (e.g. smoothed estimator) or parametric (e.g. autoregressive or periodic) model.  

\section{Detecting Variability in Light Curves}
\label{detection.sec}

\begin{figure}[b]
    \centering
    \includegraphics[scale=0.58]{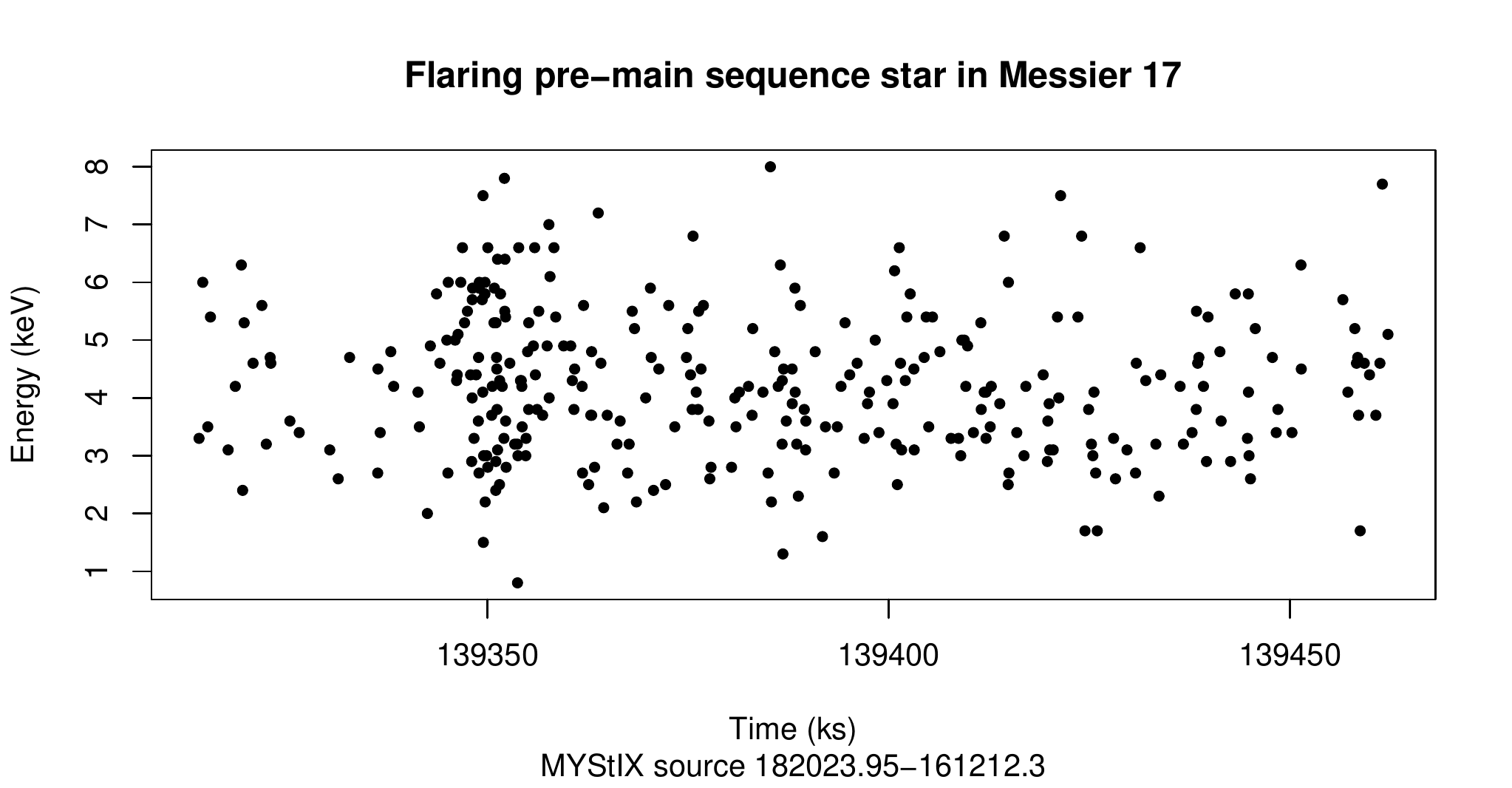}
    \caption{Time-energy diagram of 312 photons from the flaring pre-main sequence star MYStIX 182023.95-161212.3 \cite{Getman21}.}
    \label{data.fig}
\end{figure}

To illustrate many of the methods below, we use a dataset from the Chandra X-ray Observatory (Table 1, Figure~\ref{data.fig}).  Produced by a magnetically active pre-main sequence star in the Messier~17 star forming complex, the X-ray light curve has 312 photons obtained during a 148~ks observation exhibiting a prominent flare \cite{Getman21}.  Several hypothesis tests for detecting the presence of a deviation from constant emission follow in this section.

\subsection{Anderson-Darling Test}
\label{AD.sec}

\begin{figure}
    \centering
    \includegraphics[scale=0.75]{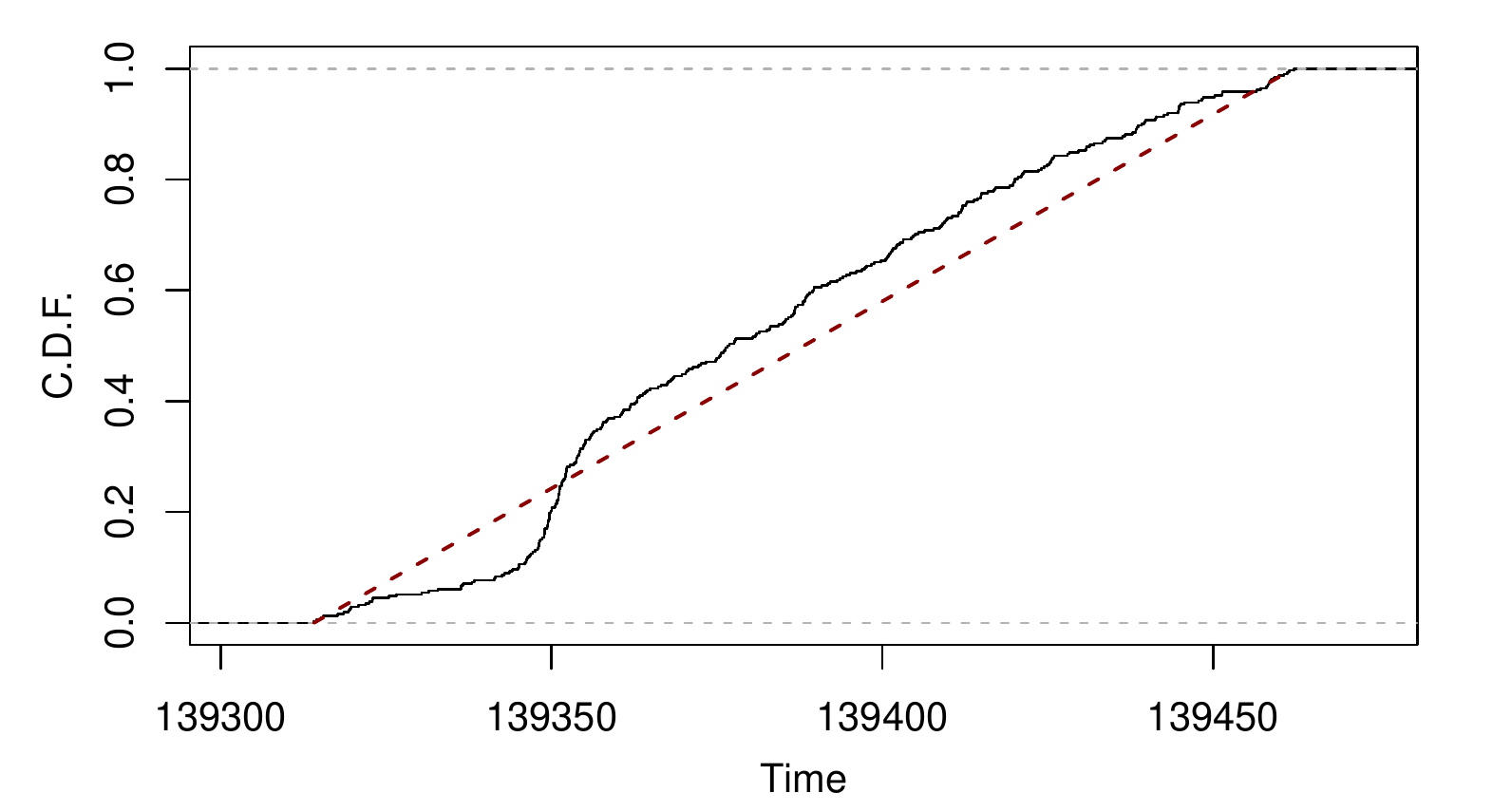}
    \caption{Cumulative distribution function of photon arrival times from Figure~\ref{data.fig} with comparison to a constant flux model (dashed line).}
    \label{ecdf.fig}
\end{figure}

Perhaps the most familiar nonparametric test for variability is the Kolmogorov-Smirnov (KS) 1-sample test based on the cumulative distribution function of photon arrival times shown in Figure~\ref{ecdf.fig}.  The KS statistic, the maximum distance between the data and the model in this graph, gives a probability $P=4$\% that the stellar X-ray emission is constant.  The KS test is not very sensitive in cases like this where the variation has a short duration.  

A better statistic is the sum of squared deviations between the data and model used in the Cramer-von Mises test, but it does not take into account the fact that all cumulative distribution functions converge at the two extremes.  The most sensitive  nonparametric statistic is the tail-weighted Cramer-von Mises statistic called the {\bf Anderson-Darling (AD) test}.  For comparison with the uniform distribution, the AD statistic is \cite{Rahman04} 
\begin{equation}
A^2 = -N - \frac{1}{N} \sum_{i=1}^N \large[ (2i-1) ln (t_i) + (2N + 1 - 2i) ln(1 - t_i) \large], 
\end{equation}
\noindent where $t_i$ are the arrival times scaled to the interval (0,1) and $N$ is the number of photons.  For the stellar data considered here, the associated p-value is $P = 0.0004$\%, far more sensitive than the KS test.  

The AD test is implemented in several locations in the R public domain statistical software environment (e.g. CRAN packages {\it goftest, ksamples, DescTools}). The implementation in Python's {\it scipy.stats} does not accept the uniform distribution for comparison.

\subsection{Test for Overdispersion}
\label{overdispersion.sec}

\begin{figure}
    \centering
    \includegraphics[scale=0.75]{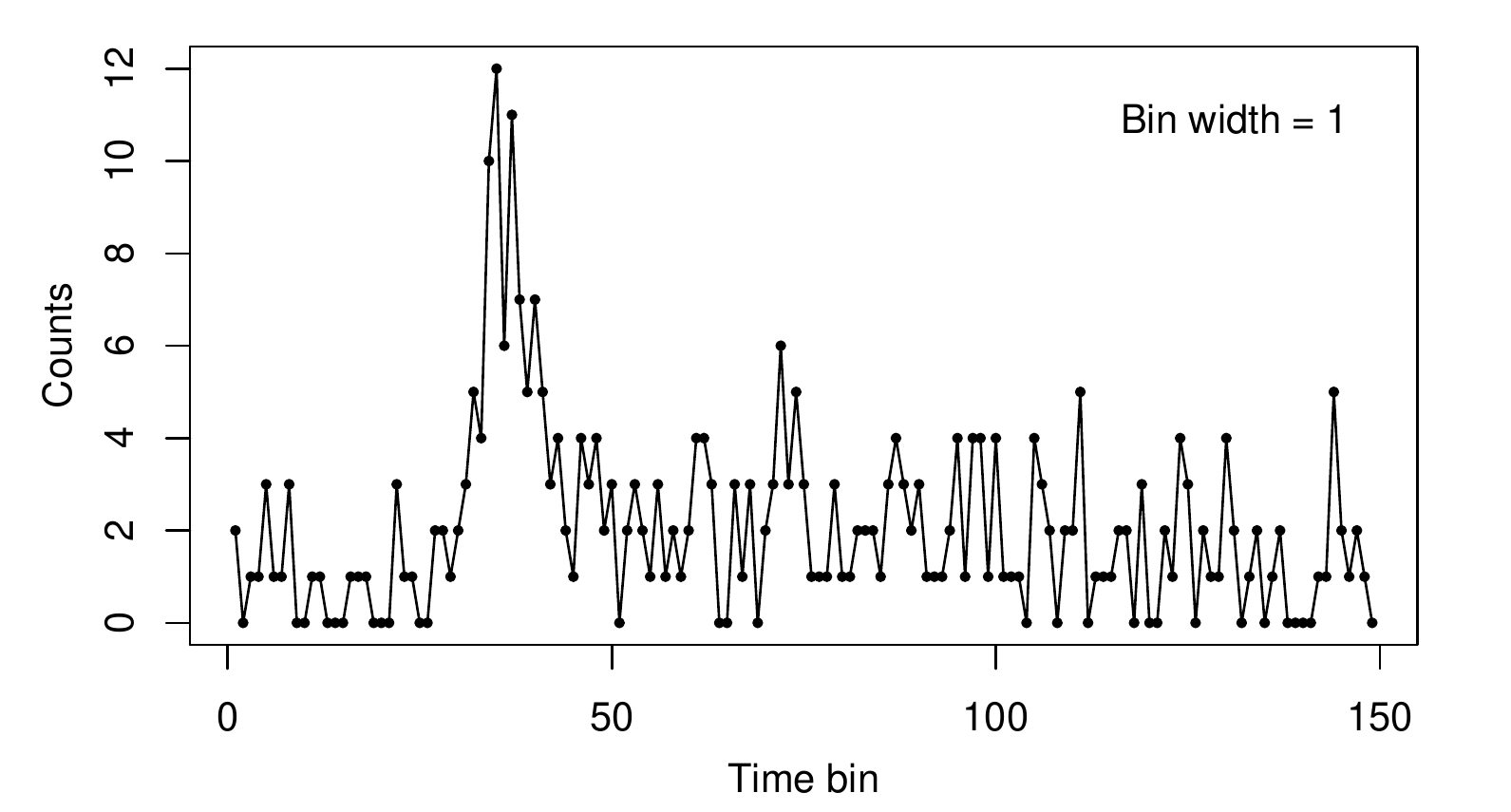} \\
    \includegraphics[scale=0.75]{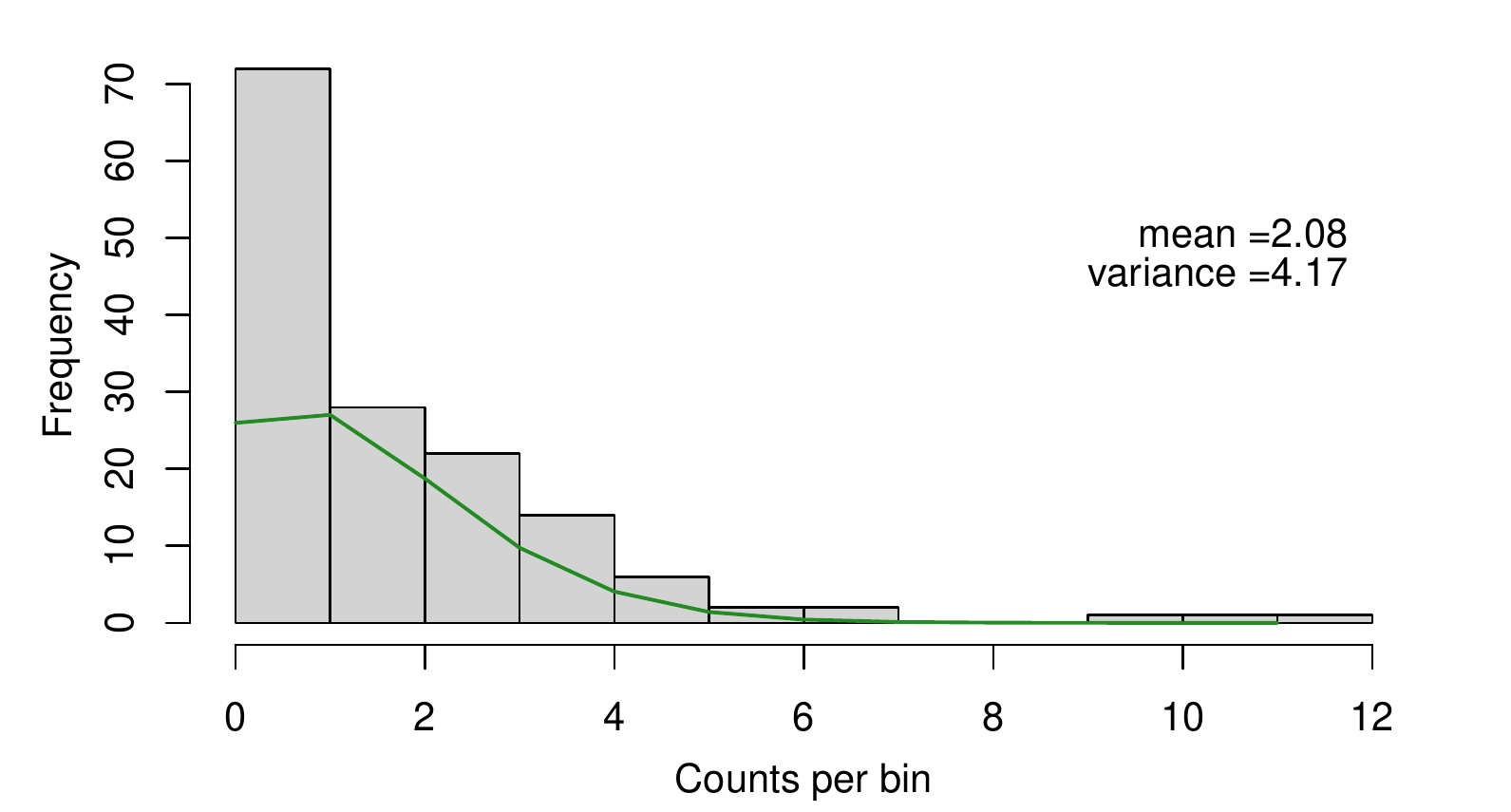}
    \caption{Top: Light curve with regularly spaced bins with 1~ks width obtained from the photon arrival times in Figure~\ref{data.fig}.  Bottom: Distribution of counts in the binned light curve.  The green curve shows the expected distribution for a homogeneous Poisson process (constant flux) with intensity 2.08.}
    \label{overdisp.fig}
\end{figure}

Though it may seem paradoxical, simple questions regarding the existence and nature of variability in an X-ray or gamma-ray source can be discerned without examining the photon arrival times.  These methods use binned data where the photons are grouped into regularly spaced intervals.  There is no need for the number of counts to be large enough so  Gaussian statistics applies; bins can have zero, one or a few counts.  This collection of binned counts, irrespective of their sequence in time, is called {\bf count data}. For the dataset shown in Figure~\ref{data.fig}, the count data for a  1 ks binwidth is shown in Figure~\ref{overdisp.fig}. 

A homogeneous (that is, constant intensity) Poisson process has the mathematical property that the mean and variance are equal.  If the variance of count data significantly exceeds the mean, then the process cannot be homogeneous and variability must be present.  This is called {\bf overdispersion} of a Poisson process.  It can arise from too many low values (called zero-inflated models) or, more commonly in astronomical light curves, too many high values. Graphically, one can compare a histogram of binned counts and the prediction of a homogeneous Poisson model (Figure~\ref{overdisp.fig}).

Formal statistical tests have been developed to detect overdispersion as described in texts like Hilbe (2014). Overdispersion tests are often applied to residuals of Poisson generalized linear models involving covariates to examine the adequacy of the models, but they can be applied to the count data with a simple model of constant intensity.  

The sample mean and variance of the photon arrival rate are 
\begin{eqnarray}
    \hat{\lambda} &=& N / N_{bin} \nonumber \\
     \widehat{Var} &=& \sum_1^n (n_i - \hat{\lambda})^2 / (N-1)
\end{eqnarray}
where $N_{bin}$ is the number of bins and $n_i$ is the number of counts in the $i^{th}$ bin.  Gaps in the data stream can be ignored under the hypothesis of constant mean flux.  The {\bf deviance goodness-of-fit test} is based on the statistic,
\begin{eqnarray}
    D &=& 2 \sum_1^{N_{bin}} {\mathcal{L}(n_i;n_i) - \mathcal{L}(\hat{\lambda};n_i)}  \\ \nonumber
   &=& \rm(Observed~variance - Theoretical~variance) \times (N_{bin}-1)
\end{eqnarray}
that is $\chi^2$-distributed with $N-1$ degrees of freedom where $\mathcal{L}$ is the likelihood. The {\bf score test} examines 
\begin{equation}
    z = \frac{\sum_1^n [(n_i - \hat{\lambda})^2 - n_i]}{\sqrt{2} \hat{\lambda}}
\end{equation}
which is $t$-distributed.  

These tests for overdispersion can be considered as a Poissonian analog of the $\chi^2$ test for variability in a Gaussian-distributed real-valued lightcurve.  Once variability has been established to be present in a Gaussian lightcurve, a common measure of the global amplitude of variation is the `excess variance', $\sigma_{xs} = S^2 - <{\sigma_{err}^2}>$ where $S^2$ is the observed variance and $\sigma_{err}$ is the measurement error of the individual measurements (e.g. Nandra et al. 1997).  However, we do not know of an analogous measure for Poisson-distributed lightcurves.  It is important not to rely on approximations of uncertainties of Poissonian variables such as Gehrels (1986) for probabilistic calculations as their statistical properties are not well-established.  

It is possible that the source may be a {\bf transient} where the counts are low except for a single outlier.  In this case, a test comparing the observed maximum counts to the maximum expected from a homogeneous Poisson process may be effective.  In the example shown in Figure~\ref{overdisp.fig}, the maximum is 12 counts in a 1~ks bin.  For a given $\hat{\lambda}$ and $n$ bins, the maximum number of counts in any bin can be predicted for a chosen significance level like $P=0.001$. However, a correction for multiple hypothesis tests must be applied, and the scientist must decide whether to apply, for example, a conservative Bonferroni family-wise correction or a more balanced Benjamini-Hochberg False Discovery Rate correction.  Due to this flexibility, we do not recomment this as a general approach to variability detection.  

Software implementation of these and related tests can be found in CRAN packages {\it COUNT}, {\it qcc}, and  {\it AER}.  A test for maximum counts in a bin is implemented in package {\it EnvStats} with a  Bonferroni correction.  More elaborate dispersion tests where the counts may be functions of covariates in generalized linear models are given in package {\it DHARMa}. No overdispersion tests are currently available in Python.

\subsection{Other Nonparametric Tests}
\label{other_nonpar.sec}

Classical statistical procedures are available for testing constancy in a stochastic process that might be useful for X-ray and gamma-ray astronomy, although they may not be very effective in the example shown in Figure~\ref{data.fig}.  

One is the {\bf Wald-Wolfowitz runs test} that evaluates the randomness of a sequence of $0-1$ binary values \cite{Conover99}.  These can be generated from the binned counts by assigning $1$ for bins above the mean value of 2.08 counts and assigning $0$ of the remaining bins.  Here the runs test gives $P=6$\% that accepts the null hypothesis of a constant flux model.  This test is not sensitive to the brief rise-and-fall event seen here, but might be effective for stationary autocorrelated behaviors and long-term secular trends.

The {\bf CUSUM test} (cumulative summation) can be applied to the cumulative distribution of unbinned photon arrival times shown in Figure~\ref{ecdf.fig}.  It was developed for industrial quality control applications.  It measures the difference between the sample means before and after each potential change point.  The test is sensitive to the variability in our example with $P < 0.0001$\% but the maximum CUSUM statistic occurs much later than the flare event. This test is implemented in CRAN package {\it CPAT}.

\subsection{Sequential likelihood-based tests}
\label{sequential.sec}

Based on the development of the broadly applicable likelihood ratio test by Wilks, Neyman and Pearson, Wald (1945) proposed the {\bf sequential probability ratio test (SPRT)} to test the hypothesis that a sequence of values with a known distribution changes its average value.  It is often used in manufacturing and clinical trials to set stopping rules.  The SPRT lies at the foundation of the statistical field of parametric sequential analysis and change point detection \cite{Tartakovsky20}. The threshold for identifying a sudden event must be chosen by the scientist in advance based on the desired rates and ratio of False Positive and False Negative triggers.  

This approach is proposed for Poisson distributed high energy lightcurves by B\'elanger (2013) who shows that an unbinned sequential likelihood calculation is very sensitive to transients that may be missed when the light curve is binned.  Here the likelihood is based on the fact that a stationary Poisson process will exhibit an exponential distribution of event inter-arrival times.  The field of sequential analysis may have other useful procedures for identifying variability or flaring in high energy sources.  

\subsection{Treatment of background events}

The astronomical signal from the source of interest $S$ is often mixed with an uninteresting background signal $B$ from the instrument or diffuse cosmic emission into an observed signal $O$.  In the simple case where $S = O - B$ are all binned count data, then the distribution of $S$ needed to estimate its accuracy follows a {\bf Skellam distribution}.  (It can not be a Poisson distribution because $O-B$ can be negative.)  This statistical result is only rarely recognized in the high energy astronomy community; see the discussion by Andr\'es et al. (2022).  

For unbinned events, it is often useful to treat the $O$ data directly and place background variations into the statistical model.  For the Anderson-Darling test (\S\ref{AD.sec}), the observed cumulative distribution function would then be compared to a curved rather than straight line model.  For cases where the background is measured in a large region and needs to be scaled to the observed region, more sophisticated treatments are needed.  

\section{Characterization of Variability}
\label{characterization.sec}

After the photon arrival times of an X-ray or gamma-ray have been shown to be inconsistent with a constant flux using a hypothesis test, then the variation needs to be characterized.  Several approaches are discussed in this section.  We start with two nonparametric approaches for stationary time series (autocorrelation and structure functions) and proceed with (semi-)parametric approaches that can be used for nonstationary time series.  Parametric approaches are called {\bf time series modeling} in the statistics literature.

\subsection{Autocorrelation function}
\label{ACF.sec}

The {\bf autocorrelation function (ACF)} of the regularly spaced binned time series is a useful transform that concentrates short-memory autocorrelation in a few small-lag coefficients. It measures the excess variance between values separated by lag $k$.  Like the Fourier transform and the autoregressive ARMA model, all information in the original data is present in the ACF if it is calculated to sufficiently large lag values.  

\begin{figure}
    \centering
    \includegraphics[scale=0.75]{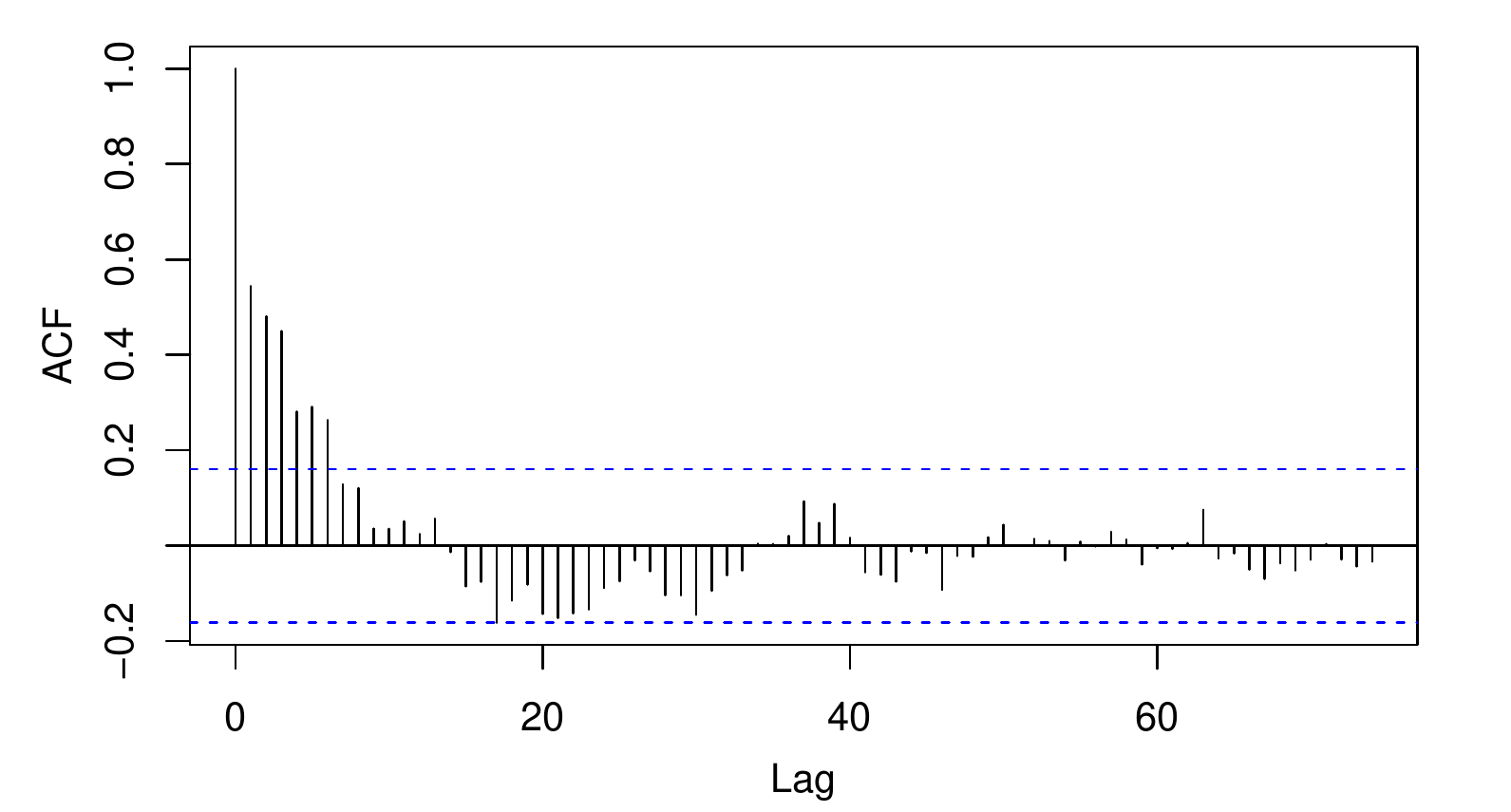}
    \caption{Autocorrelation function of binned counts from the data in Figure~\ref{data.fig}.  The dashed lines represent the 95\% confidence band for Gaussian data.}
    \label{acf.fig}
\end{figure}

The ACF has the additional advantage of having theorem-based hypothesis tests that allow p-values to be estimated for autocorrelation at $lag = 1$ (the Durbin-Watson test), $lag=k$ (the Breusch-Pagan test), and any lag up to $lag = k$ (the portmanteau Ljung-Box test).   However, the p-values for these tests are calculated in complex ways that appear to depend on Gaussian distributions \cite{Durbin71}.  We therefore doubt these confidence bands (dashed lines in Figure~\ref{acf.fig}) are accurate for integer count data.  But they still may be useful: an ACF with all values within the confidence band will clearly be consistent with uncorrelated counts, although values outside the confidence band may not definitely be inconsistent.    

The ACF for the flaring star Chandra data in Fig~\ref{acf.fig} shows a strong correlation signal at $lag=1$, $ACF(1) \simeq 0.5$, and continuing positive autocorrelation for several kilosecond bins.  This is likely dominated by the obvious flare, but there may also be contribution during the quiescent phase (see \S\ref{INAR.sec}).  

\subsection{Structure Function}
\label{SF.sec}

{\bf Structure functions (SFs)} were developed in the mid-20th century for stationary stochastic processes and have been used in certain fields of engineering and plasma physics.   It was brought into astronomical time series analysis by Simonetti et al. (1985) and have been used particularly often for characterizing high energy variability of blazars using binned data.  The widely-used 2nd-order SF, defined
\begin{equation}
    D^{(2)}(|j-k|) = \sum_j \sum_k (y_j - y_k)^2, 
\end{equation}
essentially the variance as a function of lag $j-k$.  A weighting function can be introduced for irregularly spaced observations. 

Simonetti et al. incorrectly stated that it was more sensitive to small-scale correlated variations than the autocorrelation function. Rather, the 2nd-order SF  contains exactly the same information as the ACF; they are related by an arithmetic relation \cite{Emmanoulopoulos10}. Analytic expressions for the SF are available for simple stochastic processes such as Gaussian, exponential, and power law, just as there are analytic expressions for the ACF and the Fourier power spectrum \cite{Franceschetti06}. 

The power of the SF approach is its extension to higher-orders $M>2$,  
\begin{equation}
    D^{(M)}(|j-k|) = \sum_{i=-(M-1)}^{M-1} (-1)^i 
    \begin{pmatrix}
    2(M-1) \\ M-1+i \end{pmatrix}
    ACF(i \times |j-k|).
\end{equation}
The value of higher-order SFs is illustrated in \cite{Lindsay76} who treat a complicated engineering time series with a deterministic periodicity, short-term frequency instability, and long-term $1/f$-type frequency drift.  They found that at least the 4th-order structure function was needed to characterize such behaviors. 

Emmanoulopoulos et al. (2010) criticize the overinterpretation of SFs in high energy blazar studies, even when the count rates are high enough that the Gaussian assumption holds.  They find that humps and breaks often found in the SFs are not statistically significant; they appear regularly in simulated featureless $1/f$-type red noise datasets.  They also correctly discourage fitting slopes or breaks to SFs because the points are not independent of each other.  Similarly, due to the i.i.d. assumptions underlying error analysis based on bootstrap resampling, bootstrap analysis of SF models give incorrect results.  This critique will apply to overinterpretations of the ACF as well as the 2nd-order SF, as they are essentially the same.  
 
Therefore, except when higher-order SFs are used for complicated problems with extensive data, we do not recommend use of SFs for astronomical time series analysis.  All such analysis should be made with the ACF.

\subsection{Wavelet analysis}
\label{wavelet.sec}

The behavior of nonstationary time series with complex multi-scale structure can often be elucidated with wavelet analysis. These techniques can be applied to Poisson distributed time series, though they are rarely used for faint sources in X-ray or gamma-ray astronomy. Liszka et al. (2000) produce smooth curves for a regularly spaced time series with small-N integer counts.  Bochkarev \& Belashova (2009) describe a similar technique that combines maximum likelihood estimation and wavelet thresholding.  Statistician Kolaczyk (2003) describes a broad Bayesian approach to multiscale analysis based on wavelet functions for Poisson time series.  Wavelet analysis is effective for nonstationary behaviors including flaring behavior where the events may have different timescales as well as different amplitudes and start times. 
Wavelet analysis, like Fourier analysis, is a transformation of the data and is not a parametric model.  It is therefore often difficult to establish the reliability of a chosen wavelet procedure such as a wavelet denoising threshold.

\subsection{Multiple change point model}
\label{changepoint.sec}

An important statistical model for nonstationary time series is the case of piecewise (also called `segmented') constant values where the value shifts abruptly at times called  {\bf change points}.  While statistical solutions to this problem have been long known under simplifying assumptions (e.g. single change point, constant probability of change, fluxes independent of changes), the general problem is challenging, both mathematically and algorithmically, when the times, and even the number, of change points is unknown.  The problem for unbinned Poisson processes has been addressed independently in different fields \cite{Chib98, Jackson05, Harrod06}.  

\begin{figure}
    \centering
    \includegraphics[scale=0.16]{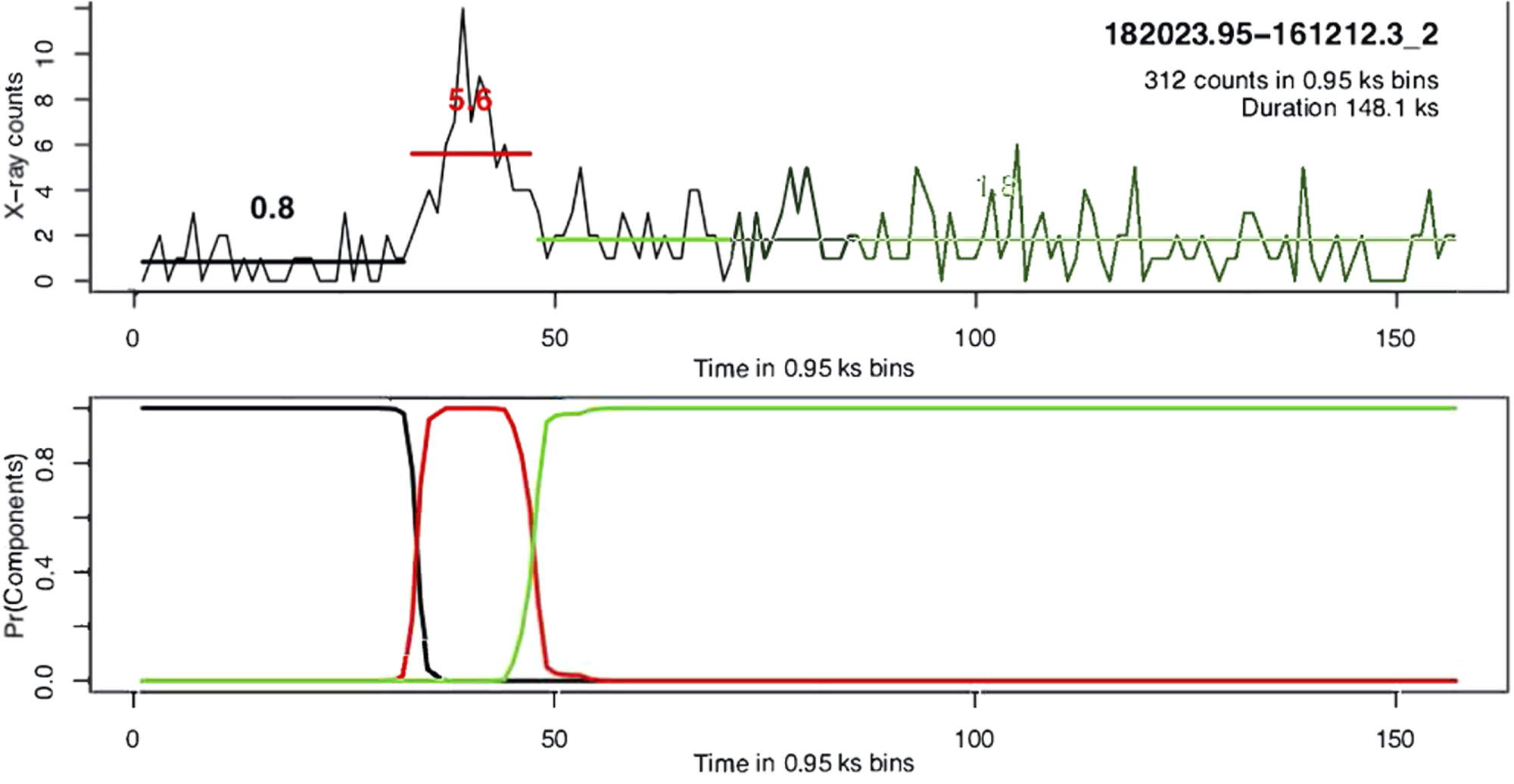}
    \caption{Multiple change point segmentation of the light curve in Figure~\ref{data.fig} using a method similar to Bayesian Blocks. The top panel shows the light curve in 0.95 ks bins with three flux levels.  The second panel shows the probabilities of each component at each time.  The scientist makes the decision rule for change points; here it chosen to be when a component exceeds 50\% probability. \cite{Getman21}}
    \label{MCMCcp.fig}
\end{figure}

In astronomy, the most important approach has been {\bf Bayesian Blocks} developed by Scargle (1998) and Scargle et al. (2013).  The likelihood is 
\begin{eqnarray}
    \mathcal{L} &=& \sum_{i=1}^{n_{cp}}{\phi(n_{cp},m_{n_{cp}}) \phi((N-n_{cp}), (M-m_{n_{cp}}))} \quad \rm{where}  \\ \nonumber
    \phi(N,M) &=& \frac{\Gamma(N+1) \Gamma(M-N+1)} {\Gamma(M+2)},
\end{eqnarray}
$N$ is the (known) number of points in the time series, $M$ is the (unknown) number of points in the current segment, and $n_{cp}$ is the (unknown) number of change points.  Uniform priors are assumed for the Bayesian calculation.  For the computation, Scargle takes a dynamic programming approach that efficiently arrives at an optimal solution with $O(N^2)$ operations and no dependency on the number of segments.  

Getman \& Feigelson (2020) use the approach of Chib (1998) and Park (2010) where the problem is formulated as a hidden Markov model and the algorithm uses Markov chain Monte Carlo techniques.  Figure~\ref{MCMCcp.fig} shows the decomposition of the lightcurve in Figure ~\ref{data.fig} into three segments using this approach.  

Code implementations of Scargle's Bayesian Block modeling is available in Matlab \cite{Scargle13}, in Python ({\it astropy.stats.bayesian\_blocks}) and in the {\it Fermitools} and {\it ISIS} packages for high energy astronomy (\S\ref{software.sec}).  The approach shown in Figure~\ref{MCMCcp.fig} is based on function {\it MCMCpoissonChange} in CRAN package {\it MCMCpack} with codes available from Brandt (2010).  

\subsection{Integer Autoregressive Models}
\label{INAR.sec}

For Gaussian distributed data that exhibit statistically signficant signals in the nonparametric ACF, low dimensional parametric models can often fit and remove the autocorrelated behavior. If the dependences are linear at different lags, the models are called ARMA (autoregressive moving average) for stationary processes and ARIMA (autoregressive integrated moving average) for nonstationary processes (e.g. with trend in the mean). Model coefficients are fit by maximum likelihood estimation and model selection, the balance between parsimonious and overfitted models, is based on the Akaike Information Criterion.  There are dozens of variants of these low-dimensional models; the GARCH model (ARIMA with volatility in variance) received the Nobel Prize in Economics.  ARIMA-type modeling has been the dominant approach to understanding stochastic time series since the 1970s \cite{Box15, Chatfield19}. 

\begin{figure}
    \centering
    \includegraphics[scale=1.00]{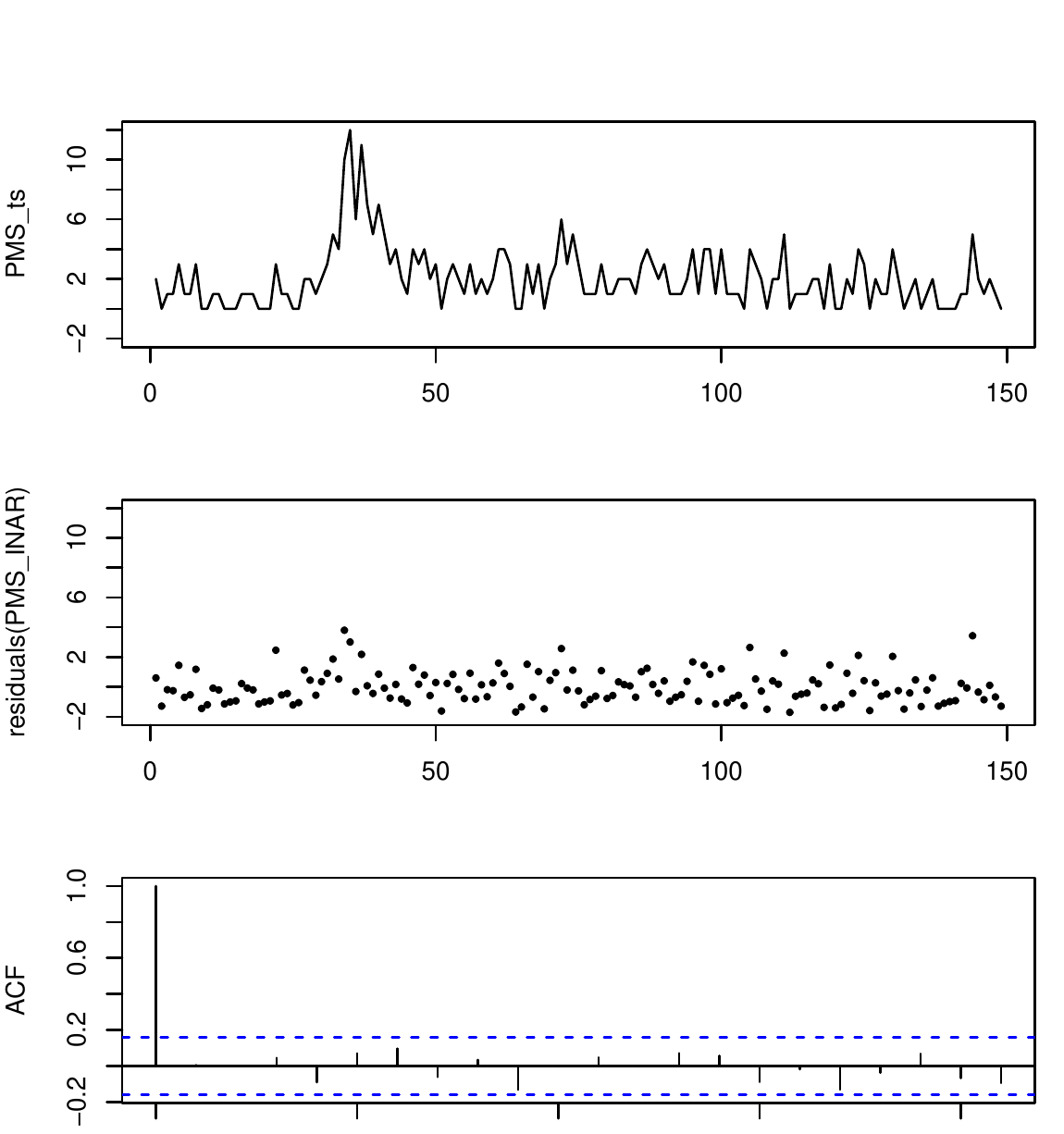}
    \caption{INAR model of the binned time series for the binned flaring star light curve.  Top: input data (same as Figure~\ref{overdisp.fig} top panel). Middle: residuals of the data and model.  Bottom: Autocorrelation function of these residuals. }
    \label{M17_INAR.fig}
\end{figure}

The standard ARIMA-type models assume time series with regularly spaced bins (although missing data may be permitted) and Gaussian errors.  However, there is a considerable literature on {\bf INteger AutoRegression (INAR)}, also called {\bf Poisson AutoRegressive (PAR)} models where the time varying quantity consists of non-negative integer counts rather than real-valued measurements.  See reviews by \cite{McKenzie03},  \cite{Jung06} and \cite{Scotto15}.  

Figure~\ref{M17_INAR.fig} shows the application of a simple model of this type, the first-order autoregressive conditional Poisson model ACP(1,1), to our X-ray stellar lightcurve.  The statistical model, also known as {\bf integer-valued generalized autoregressive conditional heteroscedasticity (INGARCH)}, is \cite{Heinen03, Jung06, 
Liboschik17},  
\begin{equation}
\mu_i  = \omega + \alpha n_{i-1} + \beta \mu_{i-1}
\end{equation}
where $\mu$ is the conditional mean or intensity of the process, $\omega$ is a Poisson-distributed random noise variable. The ACP(1,1) model is stationary and equivalent to a Gaussian ARMA process if $\alpha + \beta < 1$; the variance exceeds the  mean if $\alpha \neq 0$. The autocorrelation function for this model is
\begin{equation}
    ACF(k) = (\alpha + \beta)^{k-1} \frac{\alpha [1 - \beta(\alpha+\beta)]}{1 - (\alpha + \beta)^2 + \alpha^2}.
\end{equation}
Here the time series is treated in its entirety without any change points.   

We find in Fig~\ref{M17_INAR.fig} that this autoregressive  model with only one parameter accounts for most of the flare emission in addition to other fluctuations in the X-ray brightness.  The residuals have no autocorrelation (compare with Figure~\ref{acf.fig}) and are consistent with Poissonian white noise.

The success of this simple autoregressve model is a considerable surprise, as it implies that an appropriate stationary stochastic model can account for the entire light curve without any change points.  We suspect that if the count rate were higher, then a nonstationary model would be necessary.  It is important for high energy astronomers to test the validity of autoregressive models rather than assuming that bursts and flares are distinct events.  They may be  manifestations of a continuous autoregressive stochastic process.

Code implementation for INAR/PAR and ACP/INGARCH models are available in CRAN packages {\it acp}, {\it glarma} and {\it tscount}  while more complex models are calculated in packages {\it ConsReg}, {\it gsarima} and {\it MSwM}.  The results in Figure~\ref{M17_INAR.fig} were obtained with the {\it acp} package \cite{Vasileios15}.

\subsection{Astrophysical modeling}
\label{astrophys.sec}

Analyzing spectra with astrophysical models is essential for understanding the origins of the high energy emission but the procedure is less common for analyzing variability.  Nonetheless, it is plausible that a light curve could be modeled as a mixture of fast-rise-exponential-decay flares, or a burst of accretion from a tidal disruption event, or a model of a precessing relativistic jet.  

In such cases, model parameters are generally fit by maximum likelihood estimation with the assumption that the underlying noise process is Poissonian rather than Gaussian.  This procedure is sometimes referred to as the {\bf C-statistic} after its introduction by Cash (1979).  The statistical distribution of the C statistic is presented by Bonamente (2020) who discusses its use for parameter estimation and hypothesis testing.  Bonamente (2018) also provides formulae for evaluating the significance of faint features in binned Poissonian light curves.  

When the light curve is collected into bins with integer counts, this problem can also be considered an application of {\bf Poisson regression} where the dependence of a Poisson-distributed light curve on covariates can be estimated \cite{Hilbe14}.  In general, these are nonlinear models but they fall under the rubric {\bf generalized linear modeling} when the Poisson intensity is constant and logarithm of the flux counts expected value can be modeled by a linear combination of unknown parameters. The case of `negative binomial regression' applies when the variance and mean are not equal, as in our example here of X-rays from a flaring star.  

Poisson and related regression models are fit by maximum likelihood estimation and relevant code can be found in CRAN package {\it COUNT}.

\section{Multidimensional Variability Detection}
\label{4d.sec}

\begin{figure}
    \centering
    \includegraphics[scale=0.4]{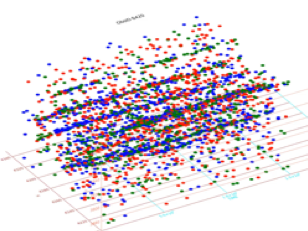}
    \caption{Four-dimensional view of a portion of the Chandra Messier 17 field  around the star MYStIX 182023.95-161212.3 shown in Figure~\ref{data.fig}. Each symbol is an X-ray photon.  The left-hand axis and the vertical axis give the celestial location, the long bottom axis gives the arrival time, and color indicates energy(red = soft, blue = hard). }
    \label{M17_4D_image.fig}
\end{figure}

As mentioned in \S\ref{foundations.sec}, many X-ray and gamma-ray instruments produce four-dimensional data with location, energy, and timing information.  Figure~\ref{M17_4D_image.fig} gives an illustration of this, showing a portion of the Chandra ACIS data around the pre-main sequence star in the Messier 17 star forming region whose light curve is shown in Figure~\ref{data.fig}.  A constant X-ray source will produce a smooth streak with uniform color.  But examination shows that many of the streaks are intermittent, indicating temporal variability, and have changing color, indicating energy variability.  This is expected because the X-ray emission from these stars is produced by powerful magnetic flares, each producing a burst of X-rays on timescales of hours.  As the flare decays, the plasma cools so that the energy spectrum evolves from higher to lower energy photons \cite{Getman21}.  The figure also show photons that are randomly distributed in location and time; this is a combination of detector background and diffuse X-ray emission from the Messier 17 nebula \cite{Townsley11}.  

Wong et al. (2016) develop a statistical approach for identifying change points in two dimensions $-$ time and energy $-$ after the photons for each star are extracted.  Nicknamed {\bf Automark}, the data are viewed as a marked Poisson process.  The distribution of photon energies is modeled using nonparametric smoothing for the continuum and a parametric lasso model for emission lines (if present).  The fitting algorithm is based on the minimum dispersion length (MDL) principle.  Models with change points in the energy and/or temporal distribution are fit by maximum likelihood estimation, and the preferred model is selected using MDL. The significance of change points is quantified with Monte Carlo simulations.  The method can find temporal variations that are restricted in energy (such as variations in emission lines but not the continuum) and it can be extended to identifying change points in the spatial dimensions.  

\begin{figure}
    \centering
    \includegraphics[scale=0.17]{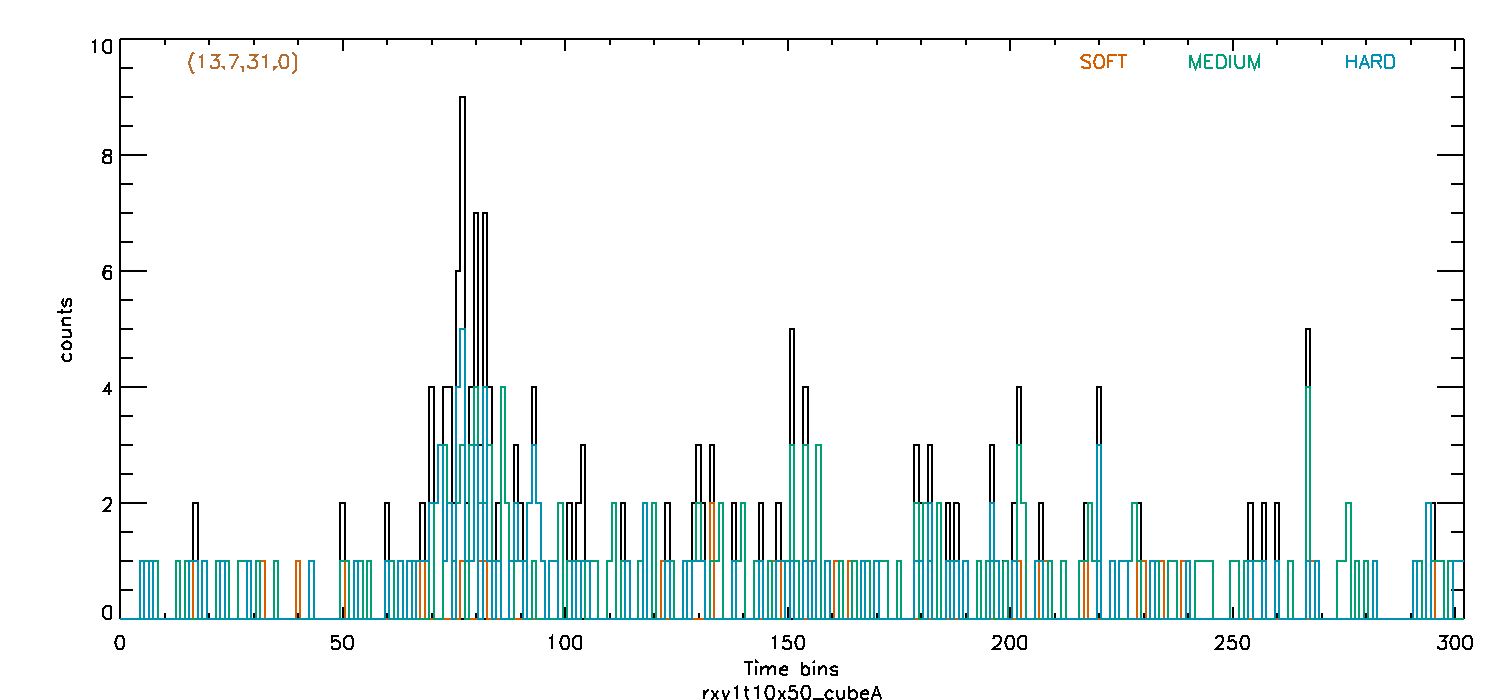} 
    \includegraphics[scale=0.17]{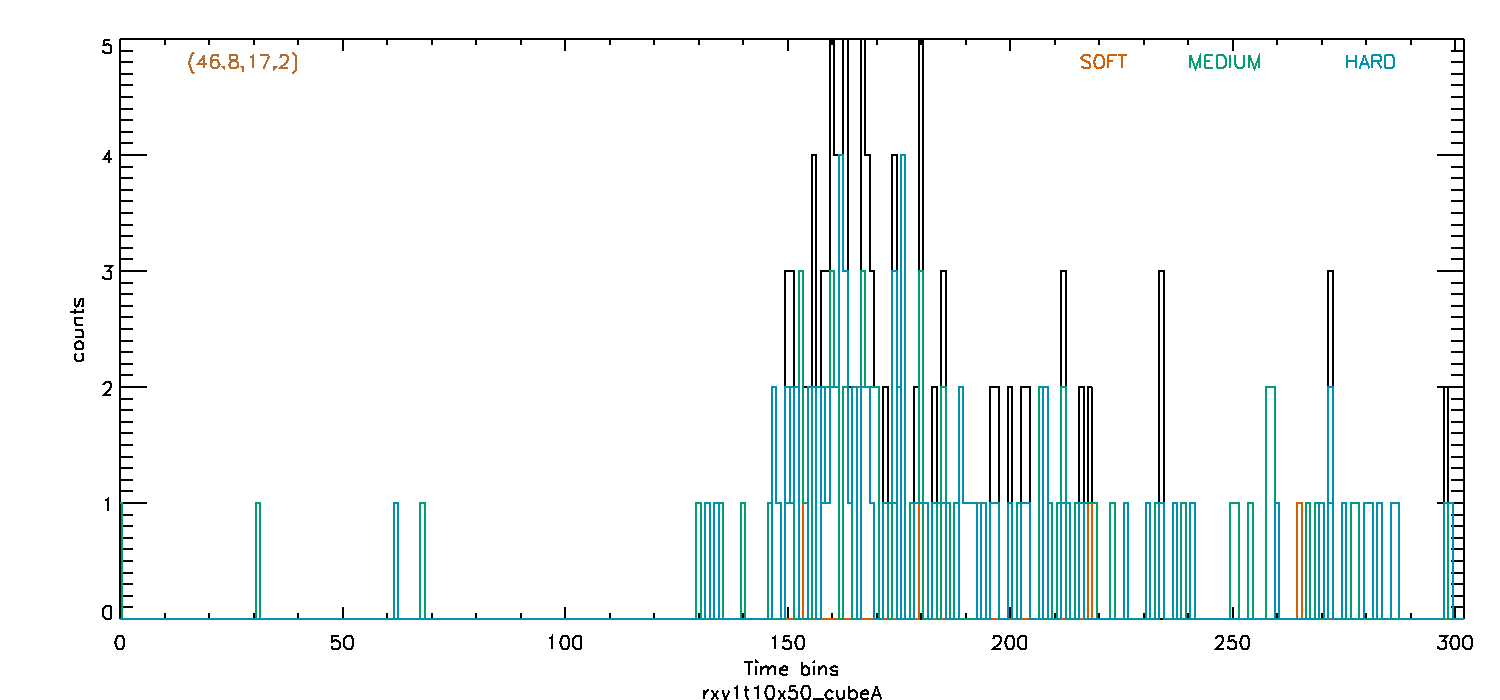}   \includegraphics[scale=0.17]{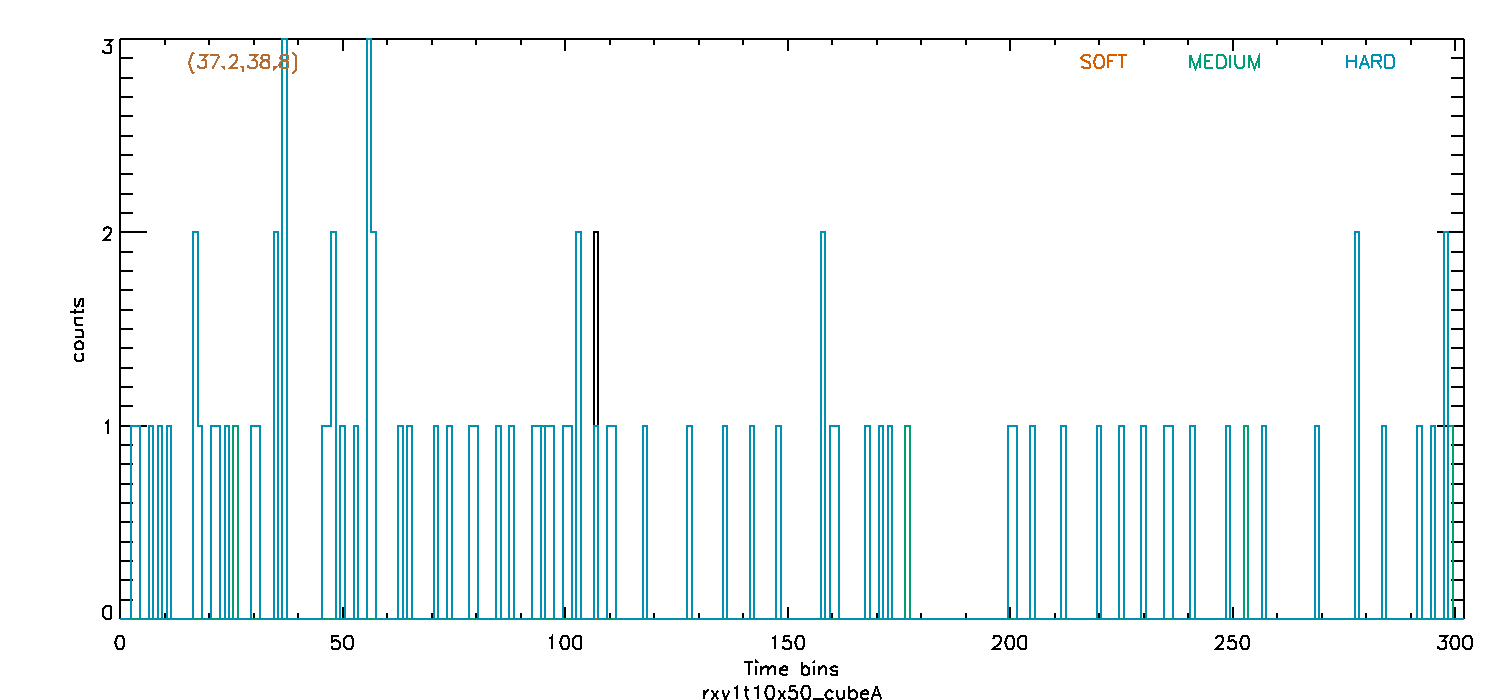}    \includegraphics[scale=0.17]{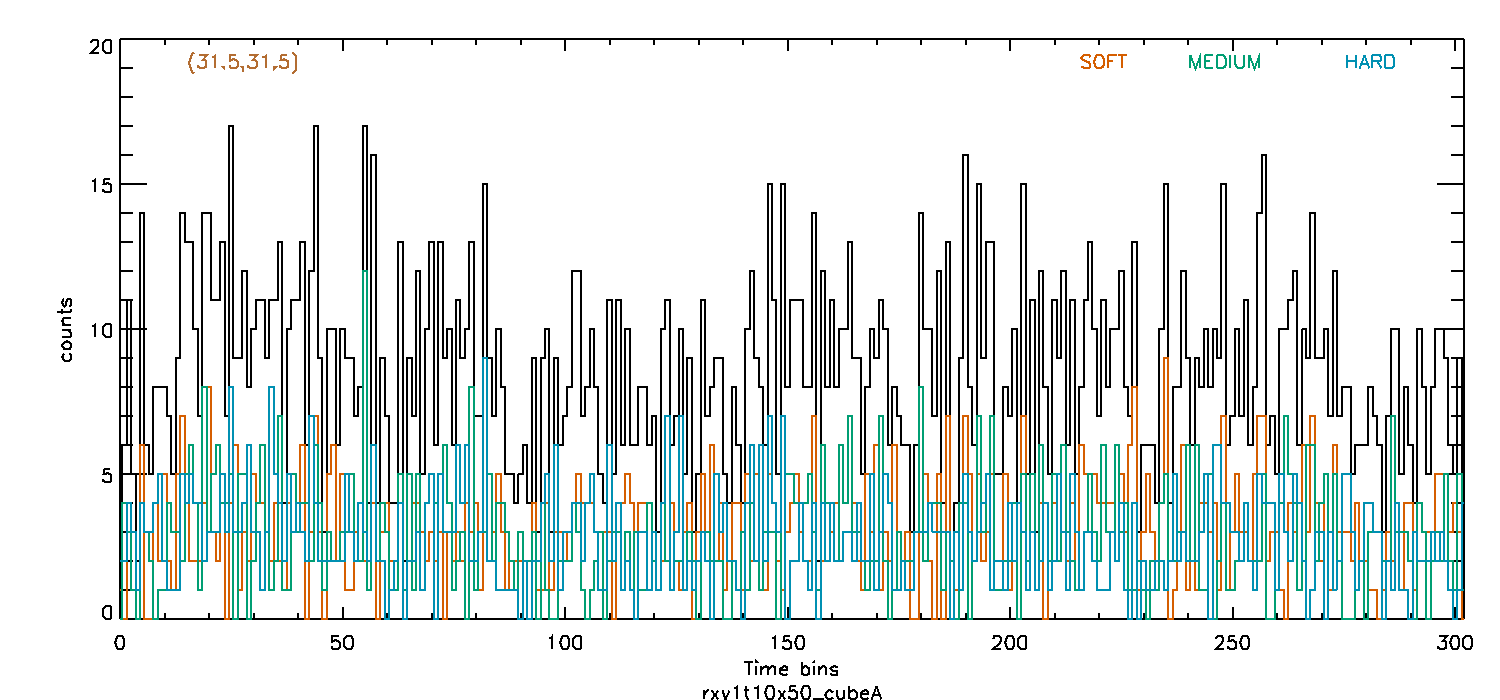}   
    \caption{Four lightcurves from the 4-dimensional analysis. From top to bottom: our target source; the most variable source; a faint variable source; and bright constant source. Soft photons are shown in orange, medium energy photons in cyan, and hard photons in black.}
    \label{M17_4D_lc.fig}
\end{figure}

Some results from this Automark analysis of the data in Figure~\ref{M17_4D_image.fig} are shown in Figure~\ref{M17_4D_lc.fig}.  The top panel shows the source MYStIX 182023.95-161212.3 that we have examined repeatedly here.  The next two panels show other variable stars, while the bottom panel shows a bright source with constant brightness and spectrum. 

Other methods treat high energy data in 3- or 4-dimensions.  If the X-ray or gamma-ray sources have spatially overlapping point spread functions, then they can be deconvolved using variability information using the {\bf eBASCS} procedure \cite{Meyer21}. 

\section{Software Packages}
\label{software.sec}

Here we briefly review the time domain analysis capabilities of various software packages designed for analysis of X-ray and gamma-ray sources.  Generally the range of methodology is slim for time domain methods, and until the packages are improved, scientists are encouraged to use more capable software environments like R and Matlab.  

The  {\it xronos} package is the principle multi-mission timing analysis toolbox for NASA, ESA and JAXA high energy missions \cite{XANADU21}.  It has strong functionalities relating to gaps in the exposure time, spectral response functions, barycentric timing corrections, and detector dead time and saturation.  For time domain methods, {\it xronos} calculates moments of the brightness distribution (mean through kurtosis), and hypothesis tests for constant flux with the Kolmogorov-Smirnov and chi-squared tests.   

{\it Fermitools} is a toolbox for analysis of data from NASA’s Fermi Gamma-ray Space Telescope \cite{Fermitools21}.  For time domain methods, it provides an implementation of Bayesian Blocks decomposition designed for prompt gamma-ray burst emission.

The {\it Chandra Interactive Analysis of Observations (CIAO)} package is designed for analysis of data from NASA’s Chandra X-ray Observatory \cite{CIAO21}.  Its thread on variability analysis does not involve statistical tests.  The associated {\it Sherpa} modeling application provides likelihood-based analysis of images and spectra, but not time series.  For testing the existence of variability, it implements the Gregory-Loredo Bayesian fitting procedure. \cite{Gregory92}.  Although originally designed for periodicity detection, the Gregory-Loredo algorithm appears to be effective for detecting a wide range of variability behaviors and taking into account low count rates \cite{Rots05}

The {\it Interactive Spectral Interpretation Systems (ISIS)} \cite{Houck00} is an analysis tool for high resolution X-ray spectroscopy from grating spectrographs and microcalorimeters. It includes an implementation of Bayesian Blocks as well as periodograms for periodicity searches. 

{\it Stingray} is a Python library designed to perform times series analysis and related tasks on astronomical light curves \cite{Huppenkothen19}. It supports a range of commonly-used Fourier analysis techniques, as well as extensions for analyzing pulsar data, simulating data sets, and statistical modeling. For time domain methods, {\it Stingray} treats multivariate time series with cross-correlation functions and estimation of time lags between similar events.  

The {\it ACIS Extract} package, written  for the imaging spectrometer on the Chandra X-ray Observatory, quantifies X-ray source variability using unequally sized time intervals chosen to give equal significance in each bin \cite{Broos10}.  

The large {\it astropy} toolbox for analysis of many types of observational astronomical data implements a False Alarm Probablity based on Scargle's Bayesian Blocks modeling, but its reliability is uncertain. 

\section{Final remarks}
\label{remarks.sec}

\begin{figure}
    \centering
    \includegraphics[scale=0.43]{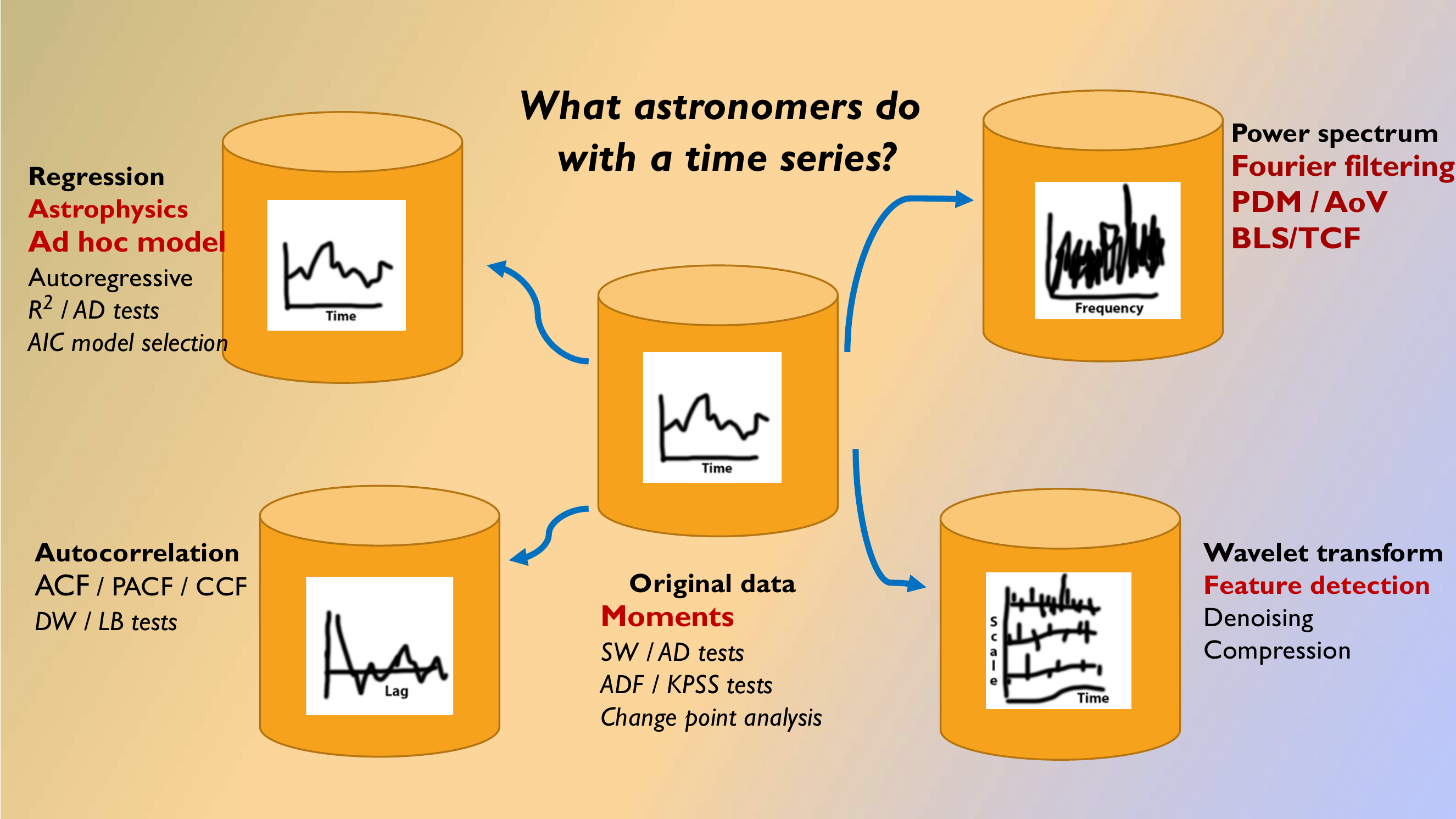} 
    \caption{Diagram of various statistical approaches to the analysis of univariate time series.  All are applicable to X-ray and gamma-ray variability studies.}
    \label{TSanal-diag.fig}
\end{figure}

From the presentation in this chapter, we see that time domain methodology for X-ray and gamma-ray astronomy in the low count regime is a non-trivial enterprise.  The diversity of temporal phenomena requires a diversity of statistical approaches, each with different strengths and limitations.  Methods designed for stationary time series may not give valid results for non-stationary time series.  Methods designed for Gaussian real-valued data may not be effective for Poisson count or event data.  

Figure~\ref{TSanal-diag.fig} illustrates some of the analysis possibilities.  When a parametric model emerges from astrophysical understanding of the emission process, such as the periodic eclipse of a binary star system, regression procedures can be pursued (upper left).  But more often for high energy variability, deterministic models are not available. Nonparametric transforms of the light curve can be useful in concentrating structure into a few coefficients to assist interpretation. These include the autocorrelation (lower left) and Fourier (upper right) transforms for stationary light curves, and the wavelet transform (lower right) for nonstationary light curves. 

The periodogram produced by the Fourier transform (and other periodograms in the frequency domain such as phase dispersion minimization and box least squares) concentrate the signal of periodic variations into a few coefficients.  But for autocorrelated behaviors like emission from a turbulent accretion disk, the signal is distributed over many coefficients and is more difficult to study.  In such cases, the nonparametric autocorrelation function and parametric ARMA-type autoregressive modeling may be more useful.  If the light curve is dominated by bursts on one or more scales, then the wavelet transform may be effective for identifying features and revealing any underlying emission behavior.    

We encourage astronomers to apply different approaches to the same dataset, recognizing that each has different strengths and weaknesses.  Unfortunately, many of the analysis procedures outlined here do not have software implementations in packages designed for X-ray and gamma-ray data analysis.  Often they are missing from the entire Pythonic environment.  Astronomers should become familiar with the R \cite{Feigelson12} and Matlab statistical software environments where a more extensive suite of analytic approaches for low count rate light curves is available.

\begin{table}
\centering
\caption{X-ray Photons from Flaring Pre-Main Sequence Star}
\begin{tabular}{lclc|lclc|lclc|lclc|lcl}
Time && E &~& Time && E &~& Time && E &~& Time && E &~& Time && E \\ \hline
139314.06 && 3.3 && 139349.89 && 3.0 && 139362.97 && 4.8 &&  139389.67 && 3.6 && 139421.02 && 5.4 \\
139314.52 && 6.0 && 139350.03 && 2.8 && 139362.98 && 3.7 && 139390.83 && 4.8 && 139421.18 && 4.0 \\
139315.15 && 3.5 && 139350.04 && 6.6 && 139363.34 && 2.8 && 139391.74 && 1.6 && 139421.41 && 7.5 \\
139315.48 && 5.4 && 139350.52 && 3.7 && 139363.80 && 7.2 && 139392.11 && 3.5 && 139423.57 && 5.4 \\
139317.68 && 3.1 && 139350.60 && 4.2 && 139364.13 && 4.6 && 139393.20 && 2.7 && 139424.03 && 6.8 \\
139318.58 && 4.2 && 139350.81 && 5.3 && 139364.48 && 2.1 && 139393.59 && 3.5 && 139424.45 && 1.7 \\
139319.33 && 6.3 && 139350.86 && 5.9 && 139364.92 && 3.7 && 139394.05 && 4.2 && 139424.90 && 3.8 \\
139319.49 && 2.4 && 139351.03 && 2.4 && 139366.14 && 3.2 && 139394.55 && 5.3 && 139425.25 && 3.2 \\
139319.68 && 5.3 && 139351.04 && 2.9 && 139366.53 && 3.6 && 139395.13 && 4.4 && 139425.43 && 3.0 \\
139320.79 && 4.6 && 139351.06 && 5.3 && 139367.46 && 2.7 && 139396.06 && 4.6 && 139425.57 && 4.1 \\
139321.89 && 5.6 && 139351.13 && 4.7 && 139367.65 && 3.2 && 139396.93 && 3.3 && 139425.78 && 2.7 \\
139322.44 && 3.2 && 139351.16 && 4.5 && 139368.03 && 5.5 && 139397.35 && 3.9 && 139425.98 && 1.7 \\
139322.92 && 4.7 && 139351.19 && 3.8 && 139368.32 && 5.2 && 139397.59 && 4.1 && 139428.01 && 3.3 \\
139323.00 && 4.6 && 139351.25 && 6.4 && 139368.56 && 2.2 && 139398.31 && 5.0 && 139428.25 && 2.6 \\
139325.38 && 3.6 && 139351.26 && 3.1 && 139369.71 && 4.0 && 139398.78 && 3.4 && 139429.68 && 3.1 \\
139326.54 && 3.4 && 139351.50 && 2.5 && 139370.30 && 5.9 && 139399.75 && 4.3 && 139430.75 && 2.7 \\
139330.35 && 3.1 && 139351.52 && 4.3 && 139370.40 && 4.7 && 139400.53 && 3.9 && 139430.84 && 4.6 \\
139331.40 && 2.6 && 139351.62 && 5.8 && 139370.69 && 2.4 && 139400.72 && 6.2 && 139431.32 && 6.6 \\
139332.82 && 4.7 && 139351.83 && 4.2 && 139371.35 && 4.5 && 139400.92 && 3.2 && 139432.02 && 4.3 \\
139336.32 && 2.7 && 139352.05 && 3.3 && 139372.19 && 2.5 && 139401.07 && 2.5 && 139433.28 && 3.2 \\
139336.35 && 4.5 && 139352.11 && 7.8 && 139372.58 && 5.6 && 139401.31 && 6.6 && 139433.67 && 2.3 \\
139336.63 && 3.4 && 139352.15 && 5.5 && 139373.34 && 3.5 && 139401.46 && 4.6 && 139433.89 && 4.4 \\
139337.95 && 4.8 && 139352.18 && 6.4 && 139374.79 && 4.7 && 139401.60 && 3.1 && 139436.30 && 4.2 \\
139338.31 && 4.2 && 139352.25 && 5.4 && 139374.96 && 5.2 && 139402.06 && 4.3 && 139436.71 && 3.2 \\
139341.34 && 4.1 && 139352.26 && 3.6 && 139375.30 && 4.4 && 139402.25 && 5.4 && 139437.79 && 3.4 \\
139341.51 && 3.5 && 139352.32 && 2.8 && 139375.50 && 3.8 && 139402.67 && 5.8 && 139438.31 && 3.8 \\
139342.51 && 2.0 && 139352.78 && 4.6 && 139375.61 && 6.8 && 139403.13 && 4.5 && 139438.34 && 5.5 \\
139342.88 && 4.9 && 139353.39 && 3.2 && 139376.01 && 4.1 && 139403.17 && 3.1 && 139438.49 && 4.6 \\
139343.65 && 5.8 && 139353.72 && 3.2 && 139376.24 && 3.8 && 139404.41 && 4.7 && 139438.64 && 4.7 \\
139344.07 && 4.6 && 139353.74 && 0.8 && 139376.31 && 5.5 && 139404.66 && 5.4 && 139439.21 && 4.2 \\
139344.93 && 5.0 && 139353.79 && 3.0 && 139376.61 && 4.5 && 139404.98 && 3.5 && 139439.59 && 2.9 \\
139345.07 && 2.7 && 139353.89 && 6.6 && 139376.90 && 5.6 && 139405.45 && 5.4 && 139439.78 && 5.4 \\
139345.11 && 6.0 && 139354.11 && 4.3 && 139377.59 && 3.6 && 139406.37 && 4.8 && 139441.26 && 4.8 \\
139345.97 && 5.0 && 139354.21 && 4.3 && 139377.70 && 2.6 && 139407.76 && 3.3 && 139441.41 && 3.6 \\
139346.15 && 4.3 && 139354.28 && 4.2 && 139377.85 && 2.8 && 139408.69 && 3.3 && 139442.59 && 2.9 \\
139346.23 && 4.4 && 139354.32 && 3.5 && 139380.41 && 2.8 && 139408.97 && 3.0 && 139443.17 && 5.8 \\
139346.33 && 5.1 && 139354.77 && 3.0 && 139380.82 && 4.0 && 139409.11 && 5.0 && 139444.69 && 3.3 \\
139346.67 && 6.0 && 139354.81 && 3.3 && 139380.96 && 3.5 && 139409.41 && 5.0 && 139444.80 && 5.8 \\ 
\end{tabular} 
\end{table}

\begin{table}
\begin{tabular}{lclc|lclc|lclc|lclc|lcl}
Time && E &~& Time && E &~& Time && E &~& Time && E &~& Time && E \\ \hline
139346.88 && 6.6 && 139355.00 && 4.8 && 139381.38 && 4.1 && 139409.60 && 4.2 && 139444.82 && 4.1 \\
139347.14 && 5.3 && 139355.13 && 3.8 && 139382.52 && 4.2 && 139409.82 && 4.9 && 139444.89 && 3.0 \\
139347.50 && 5.5 && 139355.14 && 5.3 && 139382.97 && 3.7 && 139410.55 && 3.4 && 139445.07 && 2.6 \\
139347.89 && 4.4 && 139355.72 && 4.9 && 139383.05 && 5.2 && 139411.47 && 5.3 && 139445.67 && 5.2 \\
139348.03 && 2.9 && 139355.88 && 6.6 && 139384.32 && 4.1 && 139411.52 && 3.8 && 139447.80 && 4.7 \\
139348.09 && 4.0 && 139355.97 && 4.4 && 139384.96 && 2.7 && 139411.89 && 4.1 && 139448.28 && 3.4 \\ 
139348.10 && 5.7 && 139356.18 && 3.8 && 139385.25 && 8.0 && 139412.02 && 4.1 && 139448.49 && 3.8 \\
139348.14 && 5.9 && 139356.39 && 5.5 && 139385.38 && 2.2 && 139412.12 && 3.3 && 139450.24 && 3.4 \\
139348.32 && 3.3 && 139356.87 && 3.7 && 139385.79 && 4.8 && 139412.12 && 4.1 && 139451.36 && 6.3 \\
139348.59 && 4.4 && 139357.43 && 4.9 && 139386.22 && 4.2 && 139412.64 && 3.5 && 139451.40 && 4.5 \\
139348.86 && 3.6 && 139357.65 && 7.0 && 139386.49 && 6.3 && 139412.77 && 4.2 && 139456.58 && 5.7 \\
139348.89 && 4.7 && 139357.69 && 4.0 && 139386.71 && 3.2 && 139413.83 && 3.9 && 139457.21 && 4.1 \\
139348.94 && 2.7 && 139357.79 && 6.1 && 139386.73 && 4.3 && 139414.39 && 6.8 && 139458.07 && 5.2 \\
139348.97 && 5.9 && 139358.28 && 6.6 && 139386.81 && 1.3 && 139414.88 && 2.5 && 139458.28 && 4.6 \\
139348.98 && 6.0 && 139358.50 && 5.4 && 139386.89 && 4.5 && 139414.91 && 6.0 && 139458.43 && 4.7 \\
139349.34 && 5.7 && 139359.49 && 4.9 && 139387.23 && 3.6 && 139414.97 && 2.7 && 139458.54 && 3.7 \\
139349.41 && 4.1 && 139360.37 && 4.9 && 139387.96 && 4.5 && 139415.95 && 3.4 && 139458.74 && 1.7 \\
139349.44 && 7.5 && 139360.57 && 4.3 && 139387.97 && 3.9 && 139416.85 && 3.0 && 139459.25 && 4.6 \\
139349.48 && 1.5 && 139360.77 && 3.8 && 139388.31 && 5.9 && 139417.08 && 4.2 && 139459.90 && 4.4 \\
139349.52 && 3.0 && 139360.86 && 4.5 && 139388.34 && 4.1 && 139419.24 && 4.4 && 139460.69 && 3.7 \\
139349.65 && 5.8 && 139361.79 && 4.2 && 139388.53 && 3.2 && 139419.76 && 2.9 && 139461.20 && 4.6 \\
139349.68 && 5.8 && 139361.84 && 2.7 && 139388.73 && 2.3 && 139419.82 && 3.6 && 139461.50 && 7.7 \\
139349.69 && 6.0 && 139361.97 && 5.6 && 139388.98 && 5.6 && 139419.98 && 3.9 && 139462.20 && 5.1 \\
139349.71 && 2.2 && 139362.60 && 2.5 && 139389.47 && 3.8 && 139420.06 && 3.1 &&           &&     \\
          &&     && 139362.87 && 3.7 && 139389.64 && 3.1 && 139420.49 && 3.1 &&           &&     \\

\end{tabular}
\end{table}


\vspace*{-0.2in}
\begin{thebibliography}{99}

\bibitem[Andr\'es et al. 2022]{Andres22} Andr\'es, A., van den Eijnden, J., Degenaar, N. et al.\ 2022.\ A Swift study of long-term changes in the X-ray flaring properties of Sagittarius A, {\it Monthly Notices of the Royal Astronomical Society}, 510, 2851-2863.\ doi: 10.1093/mnras/stab3407 

\bibitem[B\'elanger 2013]{Belanger13} B\'elanger, G.\ 2013.\ On detecting transient phenomena, {\it Astrophysical Journal}, 773, \#66. doi:10.1088/0004-637X/773/1/66

\bibitem[Bellm 2021]{Bellm21}. Bellm, E.~C., 2021, Review of time series features, Vera C. Rubin Observatory DMTN-118, https://dmtn-118.lsst.io

\bibitem[Bochkarev \& Belashova 2009]{Bochkarev09} Bochkarev, V.~V., Belashova, I.~A.\ 2016.\ Modelling of nonlinear filtering Poisson time series.\ {\it Journal of Physics Conference Series}, 738, 012082. doi:10.1088/1742-6596/738/1/012082

\bibitem[Bonamente 2018]{Bonamente18} Bonamente, M.\ 2018.\ Probability models of chance fluctuations in spectra of astronomical sources with applications to X-ray absorption lines.\ {\it Journal of Applied Statistics}, 46, 1129. doi:10.1080/02664763.2018.1531976

\bibitem[Bonamente 2020]{Bonamente20} Bonamente, M.\ 2020.\ Distribution of the C statistic with applications to the sample mean of Poisson data.\ {\it Journal of Applied Statistics}, 47, 2044. doi:10.1080/02664763.2019.1704703

\bibitem[Box et al. 2015]{Box15} Box, G.E.P., Jenkins, G.~M., Reinsel, G.~C. \& Ljung, G.~M., 2015, {\it Time Series Analysis: Forecasting and Control}, 5th ed., Wiley. ISBN: 9781118675021

\bibitem[Brandt 2010]{Brandt10} Brandt, P. T. 2010, Changepoint Models for Event Counts, Slides and Codes, https://personal.utdallas.edu/pxb054000/code/count-examples

\bibitem[Broos et al. 2010]{Broos10} Broos, P.~S., Townsley, L.~K., Feigelson, E.~D., Getman, K.~V., Bauer, F.~E., Garmire, G.~P.\ 2010.\ Innovations in the Analysis of Chandra-ACIS Observations.\ {\it Astrophysical Journal} 714, 1582–1605. doi:10.1088/0004-637X/714/2/1582

\bibitem[Brown et al. 2001]{Brown01} Brown L.~D., Cai, T.,  DasGupta, A.\ 2001.\ Interval Estimation for a Binomial Proportion, {\it Statistical Science},  16, 101 - 133. doi:10.1214/ss/1009213286

\bibitem[Cash 1979]{Cash79} Cash, W.\ 1979.\ Parameter estimation in astronomy through application of the likelihood ratio.\ {\it Astrophysical Journal}, 288, 939. doi:10.1086/156922

\bibitem[Chatfield \& Xing 2019]{Chatfield19} Chatfield, C. \& Xing, H., 2019, {\it The Analysis of Time Series: An Introduction with R}, 7th ed., Taylor \& Francis/CRC. doi:10.1201/9781351259446

\bibitem[Chib 1998]{Chib98} Chib, S.\ Estimation and comparison of multiple change-point models, {\it Journal of Econometrics}, 86, 221. doi:10.1016/S0304-4076(97)00115-2 

\bibitem[CIAO 2021]{CIAO21} Chandra Interactive Analysis of Observations, Chandra X-ray Center, https://cxc.cfa.harvard.edu/ciao/

\bibitem[Conover 1999]{Conover99} Conover, W. J., 1999, {\it Practical Nonparametric Statistics}, 3rd ed., Wiley. ISBN: 978-0-471-16068-7

\bibitem[Durbin \& Watson 1971]{Durbin71} Durbin, J. \& Watson, G. S., Testing for Serial Correlation in Least Squares Regression. III, {\it Biometrika}, 58, 1.  https://www.jstor.org/stable/2334313

\bibitem[Emmanoulopoulos et al. 2010]{Emmanoulopoulos10} Emmanoulopoulos, D., McHardy, I.~M., Uttley, P.\ 2010.\ On the use of structure functions to study blazar variability: caveats and problems.\ {\it Monthly Notices of the Royal Astronomical Society}, 404, 931–946. doi:10.1111/j.1365-2966.2010.16328.x

\bibitem[Enders 2014]{Enders14} Enders, W., 2014, {\it Applied Econometric Time Series}, 4th ed., Wiley. ISBN: 9781118808566

\bibitem[Feigelson \& Babu 2012]{Feigelson12} Feigelson, E.~D. \& Babu, G.~J., {\it Modern Statistical Methods for Astronomy with R Applications}, Cambridge Univ. Press

\bibitem[Fender and Belloni 2004 ]{Fender04} Fender, R., Belloni, T.\ 2004.\ GRS 1915+105 and the Disc-Jet Coupling in Accreting Black Hole Systems.\ {\it Annual Reviews of Astronomy and Astrophysics},  42, 317–364. doi:10.1146/annurev.astro.42.053102.134031

\bibitem[Fermitools 2021]{Fermitools21} Fermitools: Data Analysis for the Fermi mission, NASA High Energy Astrophysics Science Archive Research Center, https://fermi.gsfc.nasa.gov/ssc/data/analysis/

\bibitem[Franceschetti \& Riccio 2006]{Franceschetti06} Franceschetti, G. \& Riccio, D. 2006 {\it Scattering, natural surfaces, and fractals}, Academic Press. ISBN: 9780122656552

\bibitem[Gehrels 1986]{Gehrels86} Gehrels, N.\ 1986.\ Confidence limits for small numbers of events in astrophysical data, {\it Astrophysical Journal}, 303, 336. doi:10.1086/164079

\bibitem[Getman \& Feigelson 2021]{Getman21} Getman, K. V. \& Feigelson, E.~D., 2021, X-ray superflares from pre-main sequence stars: Flare energetics and frequency, {\it Astrophysical Journal}, 916, 32. doi:10.3847/1538-4357/ac00be

\bibitem[Gregory \& Loredo 1992]{Gregory92} Gregory, P. C. ;  Loredo, T. J., A new method for the detection of a periodic signal of unknown shape and period, {\it Astrophysical Journal}, 398, 146. doi:10.1086/171844

\bibitem [Harezlak et al. 2018]{Harezlak18} Harezlak, J., Ruppert, D. \& Wand, M.~P., 2018, {\it Semiparametric Regression with R}, Springer. doi:10.1007/978-1-4939-8853-2

\bibitem[Heinen 2003]{Heinen03} Heinen, A., 2003, Modelling time series count data: An autoregressive conditional Poisson model, {\it SSRN Electronic Journal}. doi:10.2139/ssrn.1117187

\bibitem[Hilbe 2014]{Hilbe14} Hilbe, J. M., 2014, {\it Modeling Count Data}, Cambridge Univ. Press.  ISBN: 9781107028333

\bibitem[Harrod \& Kelton 2006]{Harrod06} Harrod, S., Kelton, W.~D.\ Numerical methods for realizing nonstationary Poisson processes with piecewise-constant instantaneous-rate functions.\ {\it SIMULATION}, 82, 147. doi:10.1177/0037549706065514

\bibitem[Houck and Denicola 2000]{Houck00} Houck, J.~C., Denicola, L.~A.\ 2000.\ ISIS: An Interactive Spectral Interpretation System for High Resolution X-Ray Spectroscopy.\ {\it Astronomical Data Analysis Software and Systems IX}, ASP Conf. Ser. 216, 591. https://space.mit.edu/CXC/isis/

\bibitem[Huppenkothen et al. 2017]{Huppenkothen17} Huppenkothen, D., Heil, L.~M., Hogg, D.~W., Mueller, A.\ 2017.\ Using machine learning to explore the long-term evolution of GRS 1915+105.\ {\it Monthly Notices of the Royal Astronomical Society}, 466, 2364–2377. doi:10.1093/mnras/stw3190

\bibitem[Huppenkothen et al. 2019 ]{Huppenkothen19} Huppenkothen, D. et al.  2019.\ Stingray: A modern Python library for spectral timing.\ {\it Astrophysical Journal},  881. doi:10.3847/1538-4357/ab258d 

\bibitem[Jackson et al. 2005]{Jackson05} Jackson, B., Scargle, J.~D., et al.\ 2005, An algorithm for optimal partitioning of data on an interval, {\it IEEE Signal Processing Letters}, 12, 105. doi:10.1109/LSP.2001.838216

\bibitem[Jung et al. 2006]{Jung06} Jung, R. C, Kukuk, M. \& Liesenfeld, R., 2006, Time series of count data: modeling, estimation and diagnostics, {\it Computational Statistics \& Data Analysis},  51, 2350.  doi:10.1016/j.csda.2006.08.001

\bibitem[Kelly et al. 2009]{Kelly09} Kelly, B.~C., Bechtold, J., \& Siemiginowska, A.\ 2009.\ Are the variations in quasar optical flux driven by thermal fluctuations? {\it Astrophysical Journal}, 698, 895. doi:10.1088/0004-637X/698/1/895

\bibitem[Kelly et al. 2011]{Kelly2011} Kelly, B.~C., Sobolewska, M., Siemiginowska, A.\ 2011.\ A Stochastic Model for the Luminosity Fluctuations of Accreting Black Holes.{\it The Astrophysical Journal},730, 52. doi:10.1088/0004-637X/730/1/52

\bibitem[Kelly et al. 2014 ]{Kelly14} Kelly, B.~C., Becker, A.~C., Sobolewska, M., Siemiginowska, A., Uttley, P.\ 2014.\ Flexible and scalable methods for Quantifying Stochastic Variability in the Era of Massive time-domain astronomical data sets.\ {\it Astrophysical Journal} 788. doi:10.1088/0004-637X/788/1/33

\bibitem[Koen \& Lombard 1993]{Koen93} Koen, C.,  Lombard, F.\ 1993.\ The analysis of indexed astronomical time series - I. Basis methods. {\it Monthly Notices of the Royal Astronomical Society}, 263, 287, doi:10.1093/mnras/263.2.287

\bibitem[Kolaczyk 2003]{Kolaczyk03}  Kolaczyk, E.D.\ Bayesian multiscale methods for Poisson count data.\ In {\it Statistical Challenges in Modern Astronomy III}, Babu and Feigelson (eds.), Springer-Verlag, p. 89.  ISBN 0-387-95546-1

\bibitem[Liboschik et al. 2017]{Liboschik17} Liboschik, T., Fokianos, K. and Fried, R. (2017). tscount: An R package for analysis of count time series following generalized linear models. Journal of Statistical Software 82(5), 1–51, doi: 10.18637/jss.v082.i05.

\bibitem[Lindsay \& Chie 1976 ]{Lindsay76} Lindsey, W. C. \& Chie, C. M., 1976, Theory of oscillator instability based upon structure functions, in {\it Proc. IEEE}, 64:12, 1652-1666, doi:10.1109/PROC.1976.10408.

\bibitem[Lizka et al. 2000]{Lizka00} Liszka, L., Pacholczyk, A. G., Stoeger, W. R.\ 2000.\ Active Galactic Nuclei. VI. ROSAT Variability of Seyfert Galaxies. {\it Astrophysical Journal}, 540, 122.  doi:10.1086/309304 

\bibitem[Mallat 2008]{Mallat08} Mallat, S., 2008, {\it A Wavelet Tour of Signal Processing: The Sparse Way}, Academic Press. doi:0.1016/B978-0-12-374370-1.X0001-8

\bibitem[McKenzie, E. 2003]{McKenzie03} McKenzie, E., 2003, Discrete variate time series, in {\it Stochastic Processes: Modelling and Simulation}, (D.N. Shanbhag \& C.R. Rao, eds.),  Handbook of Statistics vol 21, 573, Elsevier. doi:10.1016/S0169-7161(03)21018-X

\bibitem[Meyer et al. 2021]{Meyer21} Meyer, A.~D., van Dyk, D.~A., Kashyap, V.~L.,  Campos, L.~F., Jones, D.~E., Siemiginowska, A.,  Zezas, A.\ 2021. {\it Monthly Notices of the Royal Astronomical Society}, 506, 6160. doi:10.1093/mnras/stab1456

\bibitem[Nason 2008]{Nason08} Nason, G., 2008, {\it Wavelet Methods in Statistics with R}, Springer

\bibitem[Nandra et al. 1997]{Nandra97} Nandra, K.,  George, I. M.,  Mushotzky, R. F., Turner, T. J., Yaqoob, T.\ 1997.\ ASCA Observations of Seyfert 1 Galaxies. I. Data Analysis, Imaging, and Timing, {\it Astrophysical Journal}, 476, 70. doi:10.1086/303600 

\bibitem[Park et al. 2006]{Park06} Park, T.,  Kashyap, V.~L.,  Siemiginowska, A., van Dyk, D.~A.,  Zezas, A., Heinke, C., Wargelin, B.~J.\ 2006.\ Bayesian estimation of hardness ratios: Modeling and computations, {\it Astrophysical Journal}, 652, 610-62. doi:10.1086/507406

\bibitem[Park 2010]{Park10}  Park, J.~H.\ 2010. Structural Change in U.S. presidents' use of force.\ {\it American Journal of Political Science}, 54, 766. https://www.jstor.org/stable/27821951

\bibitem[Priestley 1981]{Priestley81} Priestley, M.~B., 1981, {\it The Spectral Analysis and Time Series}, 2 vols. Academic Press

\bibitem[Rahman \& Chakrobartty 2004 ]{Rahman04} Rahman, M., Chakrobartty, S., 2004, Tests for uniformity : A comparative study, {\it Journal of Korean Data \& Information Science Society}, 15, 211, https://www.koreascience.or.kr/article/JAKO200423421079798.page

\bibitem[Rots 2005]{Rots05} Rots, A., Effectiveness of the Gregory-Loredo algorithm for detecting temporal variability in Chandra data, Chandra Science Center memo, https://cxc.harvard.edu/csc/memos/files/Rots\_GLvary2.pdf

\bibitem[Ryan et al. 2019]{Ryan2019} Ryan, J.~L., Siemiginowska, A., Sobolewska, M.~A., Grindlay, J.\ 2019.\ Characteristic Variability Timescales in the Gamma-Ray Power Spectra of Blazars.\ {\it The Astrophysical Journal} 885. doi:10.3847/1538-4357/ab426a

\bibitem[Salles et al. 2019]{Salles19} Salles, R., Belloze, K., Porto, F., Gonzalez, P., Ogsasawara, E.\ 2019.\  Nonstationary time series transformation methods: An experimental review, {\it Knowledge-Based Systems}, 164, 274. doi:10.1016/j.knosys.2018.10.041

\bibitem[Scargle 1981]{Scargle81} Scargle, J.~D.\ 1081, Studies in astronomical time series analysis. I - Modeling random processes in the time domain.\ {\it Astrophysical Journal Supplements}, 41, 1. doi:10.1086/190706

\bibitem[Scargle et al. 1993]{Scargle93} Scargle, J.~D., Steiman-Cameron, T., Young, K., Donoho, D.~L., Crutchfield, J.~P., Imamura, J.\ 1993.\ The Quasi-periodic oscillations and very low frequency noise of Scorpius X-1 as transient chaos: A dripping handrail?\ {\it Astrophysical Journal}, 411, L91. doi:10.1086/186920

\bibitem[Scargle 1998]{Scargle98} Scargle J.~D.\ Studies in astronomical time series analysis. V. Bayesian Blocks, a new method to analyze structure in photon counting data, {\it Astrophysical Journal}, 504, 405.\ doi:10.1086/306064

\bibitem[Scargle et al. 2013]{Scargle13} Scargle, J.~D., Norris, J.~P.,  Jackson, B. \&  Chiang, J., Studies in Astronomical Time Series Analysis, VI. Bayesian Block Representations, {\it Astrophysical Journal}, 764, 167. doi:10.1088/0004-637X/764/2/167

\bibitem[Scotto et al. 2015]{Scotto15} Scotto, M. G., Weiss, C. H. \& Gouveia, S., Thinning-based models in the analysis of integer-valued time series: a review, {\it Statistical Modeling}, 15, 590. doi:10.1177/1471082X15584701

\bibitem[Simonetti et al. 1985]{Simonetti85} Simonetti, J.~H., Cordes, J.~M., Heeschen, D.~S.\ 1985.\ Flicker of extragalactic radio sources at two frequencies,\ {\it Astrophysical Journal}, 296, 46–59. doi:10.1086/163418

\bibitem[Takezawa 2006]{Takezawa06} Takazawa, K. {\it Introduction to Nonparametric Regression}, Wiley. ISBN: 9780471771456

\bibitem[Tartakovsky et al. 2020]{Tartakovsky20} Tartakovsky, A., Nikiforov, I. \& Basseville, M., 2020, {\it Sequential Analysis: Hypothesis Testing and Changepoint Detection}, CRC Press. ISBN: 9780367740047

\bibitem[Townsley et al. 2011]{Townsley11} Townsley, L.~K. et al.\ 2011.\ The integrated diffuse X-ray emission of the Carina Nebula compared to other massive star-forming regions. {\it Astrophysical Journal Supplements}, 194, \#16. doi:10.1088/0067-0049/194/1/16

\bibitem[Vasileios 2015]{Vasileios15} Vasileios, S., 2015, acp: Autoregressive Conditional Poisson, R package version 2.1, https://CRAN.R-project.org/package=acp

\bibitem[Vaughan et al. 2003]{Vaughan03} Vaughan, S.,  Edelson, R.,  Warwick, R. S.,  Uttley, P.\ 2003.\ On characterizing the variability properties of X-ray light curves from active galaxies, {\it Monthly Notices of the Royal Astronomical Society}, 345, 1271. doi:10.1046/j.1365-2966.2003.07042.x

\bibitem[Wald 1945]{Wald45} Wald, A.\ Sequential tests of statistical hypotheses.\ {\it Annals of Mathematical Statistics}, 16, 117. doi:10.1214/aoms/1177731118

\bibitem[Wong et al. 2016]{Wong16} Wong, R.~K.~W., Kashyap, V.-L, Lee, T.~C.~M., van Dyk, D.~A.\ 2016.\ Detecting abrupt changes in the spectra of high-energy astrophysical sources. {\it Journal of Applied Statistics}, 10, 1107. doi:10.1214/16-AOAS933

\bibitem[Xanadu 2021]{XANADU21} Xanadu: Data analysis for X-ray astronomy, NASA High Energy Astrophysics Science Archive Research Center, https://heasarc.gsfc.nasa.gov/xanadu/xanadu.html

\end{thebibliography}
\end{document}